\documentstyle [12pt] {article}
\input{epsf}  
\begin{document}

\begin{flushright} CLNS 96/1399
\end{flushright}

\bigskip\bigskip
\bigskip

\centerline{\Large {\bf Peculiarities of Quantum Mechanics:}}
\bigskip
\centerline{\Large {\bf Origins and Meaning}}

\bigskip\bigskip\bigskip
\bigskip
\bigskip
\centerline{\large {\bf Yuri F. Orlov}}

\bigskip

\centerline{Floyd R. Newman Laboratory of Nuclear Studies}
\centerline{Cornell University}
\centerline{Ithaca, New York 14853, USA}

\bigskip\bigskip
\bigskip\bigskip
\bigskip\bigskip
\bigskip

\centerline{\bf Abstract}

{\small
\begin{quotation}

The most peculiar, specifically quantum, features of quantum
mechanics~---~quantum nonlocality, indeterminism, interference of
probabilities, quantization, wave function collapse during
measurement~---~are explained on a logical-geometrical basis.  It is
shown that truths of logical statements about numerical values of
quantum observables are quantum observables themselves and are
represented in quantum mechanics by density matrices of pure states.
Structurally, quantum mechanics is a result of applying non-Abelian
symmetries to truth operators and their eigenvectors~---~wave
functions.  Wave functions contain information about conditional truths
of all possible logical statements about physical observables and
their correlations in a given physical system.  These correlations are
logical, hence nonlocal, and exist when the system is not observed.
We analyze the physical conditions and logical and decision-making
operations involved in the phenomena of wave function collapse and
unpredictability of the results of measurements.  Consistent
explanations of the Stern-Gerlach and EPR-Bohm experiments are
presented.

\end{quotation} }

\pagebreak

\centerline{\bf CONTENTS}

\bigskip

\noindent Introduction. \dotfill\ 4

\begin{enumerate}

\item Truths of statements about observables, as observables.  The
meaning \\ of density matrix and wave function. \dotfill\ 9

\item  Representations of complex statements.  Logical  correlations
between \\ noncommuting observables. \dotfill\ 15

\item Origins of quantization.  Appearance of noncomputable
functions. \\  Quantum Hilbert space. \dotfill\ 23

\item Stern-Gerlach experiment.  Quantum nonlocality  (uncertainty).
\\ Measurement of nonlocal observables. \dotfill\ 28

\item Measurement.  Indeterminism.  Collapse of wave functions. \\
Quantum $\longrightarrow$ classical transition. \dotfill\ 35

\item EPR-Bohm experiments.  The main questions and answers. \dotfill\
41 

\end{enumerate}

\noindent Conclusion. \dotfill\ 47

\noindent Acknowledgements. \dotfill\ 47

\noindent References. \dotfill\ 48

\noindent Figures. \dotfill\ 49

\pagebreak

\centerline{\bf Introduction}

\bigskip

This paper analyzes only a small fragment of quantum mechanics,
namely, quantum mechanics without second quantization and quantum
field theory.  However, most strange quantum properties~---~quantum
nonlocality, indeterminism, interference of probabilities,
quantization, collapse of wave functions during measurements, are
exposed perfectly well in this fragment.  Our analysis is not a new
version of quantum mechanics, and not even a new interpretation; it
provides only a consistent explanation of and a logical ``picture''
for the basic ideas of the Copenhagen interpretation.

It is shown here that strange quantum properties appear because, in
quantum mechanics, non-Abelian symmetries are applied to truth
operators of logical statements about numerical values of physical
observables, while in classical mechanics, symmetries are applied to
numerical values of observables themselves.  These truth operators (as
shown in Sec.~1) are also quantum observables, nonlocal by nature,
and are represented in quantum mechanics by density matrices of pure
states.  The inherent contradiction between quantum logical nonlocality
and classical locality of measurements leads to the appearance of
noncomputable functions (see Secs.~3 and 5), and hence to indeterminism,
probabilistic behavior of quantum systems, and wave function
collapse.  This does not mean, however, that we cannot measure
nonlocal observables (as shown in Sec.~4).

The concept of probability makes sense only in connection with
measurement, since truth operators~---~density matrices~---~develop
deterministically between measurements, obeying quantum equations.

A given state vector of a physical system contains information about
the conditional truths of all possible logical statements describing
physical observables relevant to the system, provided that this system
is in the given quantum state (see Sec.~2).  And since such a state
can be destroyed by a measurement, the concept of conditional truth
applies only between measurements.  So, all possible local and
nonlocal correlations are present in the logic of the physical system
{\underline{before}} the system is observed.  This logic, by its
nature, does not need to be localized.  The physical system itself may
be placed somewhere, but in most cases a true statement about its
position is not expressible (see below).

Although, numerically, conditional truths are defined in such a way
that they are equal to the corresponding probabilities, this does not
mean that nonlocal correlations observed in measurements are results
of mutual influences of spatially separated measuring procedures.
These correlations were born together with the initial state of the
physical system.  (The problem of the origins of nonlocal correlations
has created a huge literature; we refer readers only to [1].  We
analyze the EPR-Bohm experiments (Sec.~6) in connection with this
problem.)

It follows from the foregoing that a quantum Hilbert space appears as
a result of applying symmetry transformations to a full set of
eigenvectors belonging to originally commuting truth operators of
statements describing a given physical system.  The result is a 
quantum logic of noncommuting truth operators.  In this logic,
symmetry transformations are merely redefinitions of meanings of
``yes's'' and ``no's''.  From this point of view, the strangest
quantum features are more linguistic than physical problems.
Structurally, any quantum Hilbert space is a continuum of mutually
noncommuting (though equivalent) {\underline{classical}} logics and
languages interconnected with each other by a set of conditional
truths.  By definition, conditional truths vary between 0 (``false'')
and 1 (``true'').  The conditional truth of a complex statement 
containing simpler constitutent statements makes sense only if it
does not depend on the order of those constituent statements.  It is
important that such independence exists~---~at least
under special conditions.  (This is discussed in Sec.~2.)

The question arises why logic and language play such a fundamental
role in quantum mechanics, while in classical  mechanics they play only
an auxiliary one.  The following is a qualitative explanation.

Though undescribed Nature certainly exists, scientific knowledge of
Nature exists only in the form of logically organized descriptions.
When these descriptions become ``too precise'' at some level of
accuracy, the fundamental features of logic and language acquire the
same importance as the features of what is being described.  At this
``micro''-level, we cannot separate the features of ``matter per se''
from the features of the logic and language used to describe it.  In
particular, two properties of classical mathematical logic are
potentially ``quantum'' and become crucial at the micro-level of
accuracy:
\begin{enumerate}

\item[(a)] The truth values of classical logic are naturally
quantized.

\item[(b)] There exists a hidden, unformalized symmetry in
classical logic~---~namely, any logical tautology remains the same
tautology, regardless of how we change the meaning of the truth values
of its constituent statements.  The only requirement is that every
newly redefined ``no'' be the negation of a correspondingly redefined
``yes.''

\end{enumerate}

So we immediately get a quantum-like logical system when we formalize
these properties (as done in [2]) using the assumptions that
symmetries are linear, continuous and non-Abelian.

The known features of quantum measurements are consistent with this
logical picture.  In quantum mechanics, when our measurements become
``too precise,'' i.e., when they reach a quantum scale of accuracy, we
cannot exclude the influence of measuring processes on measuring
systems. As a result, we lose the possibility of unconditionally
defining truth: the definition of truth now depends on how we observe
the physical system, on our choice of apparatus.  This logical
relativism does not exist in classical mechanics, where logical
statements are precise logically but not physically because they
describe huge intervals of quantum numerical values, $\Delta S/\hbar
\gg 1$; as a result, all truth operators of logical statements about
observables commute and truth values are never redefined
(transformed).  Not being transformed, classical truth values do not
depend on choice of apparatus.

In quantum mechanics, truths of logical statements about dynamic
variables become dynamic variables themselves, because they depend on
parameters of symmetry transformations that redefine truth values.

If the truths of statements become dynamic variables, then
{\underline{whose}} statements are they?  The answer is that abstract
sets of all possible languages and statements describing physical
observables exist objectively, as do sets of the conditional truths of
those statements, whereas the choice of a language (= quantum
representation) and questions to be answered, as well as the
formulation of statements describing the results of measurements, 
belong to the researcher.

Another question arises about whether we can verbally express
statements 
when their truths are quantum operators.  The answer is that since
statements themselves are not operators, they can be (and always
really are) expressed verbally in an ordinary way.  That, say, $p$ and
$q$ are noncommuting observables does not limit our formulations of
any statements about exact numerical values of both noncommuting $p$
and $q$.  In fact, when describing any observable informally, we
always use its own representation.  The truths of such statements,
however, cannot be expressed explicitly and simultaneously for both
$p$ and $q$.  If, for example, it is true that $p = p_0$, then nothing
precise can be said about the truth of the statement $q = q_0$.
Still, we can describe numerically the conditional truth of this
statement, under the condition $p = p_0$.

We should note that in [2] and [3], logical statements are not
distinguished from their truth, so truth operators are simultaneously
operators of statements themselves.  This representation of statements
by their truth values can be formalized (see, for example, [4], \S45);
however, in practice, such mixing can create ambiguity.  In this
paper, we avoid such mixing.

\bigskip\bigskip

\pagebreak

\noindent {\bf 1.~~Truths of statements about observables, as
observables.  The meaning of density  matrix and wave function.}

\bigskip

Let $K$ be an observable with a set of possible numerical values
(quantum numbers), $\{ k_1, k_2, \ldots \}$, and let a physical system
be in  state $\mid k_i >$.  The logical statement $\Lambda_{k_i}$,
$$
\Lambda_{k_i} {\mbox{:~~``The system is in state}}~ \mid k_i
> {\mbox{''}}~~, 
\eqno(1a)
$$
\noindent or, equivalently,
$$
\Lambda_{k_i}{\mbox{:~~ ``}} K = k_i{\mbox{~''}}~~,
\eqno(1b)
$$
\noindent describes the real situation in this case and therefore is
true. 

We will prefer to evaluate truths of statements numerically; let the
truth value ``true'' be assigned the number 1, and the truth value
``false'' the number 0.  In our case, the truth of $\Lambda_{k_i}$ is
equal to 1.

This truth value can be confirmed by measurement.  Our apparatus
should measure $K$; if after (theoretically infinitely) many
repetitions of the same experiment, we get the same number, $K=k_i$,
then the truth of $\Lambda_{k_i}$ is equal to 1, while all statements
$\Lambda_{k_j}$, $k_j \neq k_i$, are false and their truths  are equal
to zero.  Thus, the truth of $\Lambda_{k_i}$ is measured
simultaneously with $K$ and is an observable which can be called a
logical observable.  We can represent this observable by a Hermitian
``truth operator'', $\hat{\Lambda}_{k_i}$, commuting with the operator
$\hat{K}$ representing observable $K$.

Since any truth operator, $\hat{M}$, possesses only two exact numerical
values, 1 and 0, it is a projector:
$$
\hat{M}^2 = \hat{M}~~.
\eqno(2)
$$

Since all $\hat{\Lambda}_{k_i}$, $i = 1, 2, \ldots$, commute with
$\hat{K}$, the eigenvectors of $\hat{\Lambda}_{k_i}$ and $\hat{K}$ are
the same, so $\hat{\Lambda}_{k_i} \mid k_j > = \delta_{k_i k_j} \mid
k_j >$, or, in matrix form,
$$
\sum_{k^\prime} \Lambda_{k_i} (k, k^\prime) \mid k_j (k^\prime) > =
\delta_{k_i k_j} \mid k_j (k) >~;~~ \delta_{k_i k_j} = \left\{
\begin{array}{ll} 1, & k_j = k_i \\ 0, & k_j \neq k_i \end{array} 
\right. ~~.
\eqno(3a)
$$
\noindent Here $\delta_{k_i k_j}$ is an eigenvalue of
$\hat{\Lambda}_{k_i}$.  It follows from (3a) that for any $\mid \psi >
= \sum a_j \mid k_j >$, 
$$
\sum_i \Lambda_{k_i} \mid \psi > = \mid \psi >~~.
\eqno(3b)
$$
\noindent In the diagonal $K$-representation, $\mid k_j (k) > =
\delta_{kk_j}$; using this in (3a) we get $\hat{\Lambda}_{k_i}$ in
the diagonal $K$-representation:
$$
\Lambda_{k_i} (k, k^\prime) = \delta_{kk_i} \delta_{k^\prime k_i}~~; 
\eqno(4)
$$
\noindent in this representation, all matrix elements of
$\hat{\Lambda}_{k_i}$  equal  0 except for a single 1 at the $i$-place
on the main diagonal.  This means that the truth operators are density
matrices of pure states; in the diagonal representation,
$$
\Lambda_{k_i} (k, k^\prime) = \mid k_i (k) ><k_i (k^\prime) \mid~~. 
\eqno(5)
$$
 
From (4) and (5), we get:
$$
tr \hat{\Lambda}_{k_i}  =  1~~, 
\eqno(6)
$$

$$
\begin{array}{rcl}
\hat{\Lambda}_{k_i} \hat{\Lambda}_{k_j} & = & \sum_{k^\prime} \mid k_i
(k)><k_i (k^\prime) \mid k_j (k^\prime)><k_j (k^{\prime \prime}) \mid
= \\
 & = & \delta_{k_i k_j} \hat{\Lambda}_{k_i}~~,
\end{array}
\eqno(7)
$$
\noindent  where $\hat{\Lambda}_{k_i} \hat{\Lambda}_{k_j}$ is the
matrix product.  And
$$
\hat{K} = \sum_i k_i \hat{\Lambda}_{k_i}~~.
\eqno(8)
$$

In Sec.~2  we will investigate complex statements composed of 
elementary statements represented in quantum mechanics by density
matrices of pure states.  Equation (6) is valid only for elementary
statements.  Equation (7) describes the features of logical
conjunction, ``$\Lambda_{k_i}$ and $\Lambda_{k_j}$'', composed of two
elementary statements.  Repetition of the same statement,
$\Lambda_{k_i}$, i.e., ``$\Lambda_{k_i}$ and $\Lambda_{k_i}$'' is
logically equal to the same statement; and in such a case
$\delta_{k_i k_i} = 1$, $\hat{\Lambda}_{k_i} \hat{\Lambda}_{k_i} =
\hat{\Lambda}^2_{k_i}$, and (7) coincides with (2).  If $k_i \neq
k_j$, then $\Lambda_{k_i}$ and $\Lambda_{k_j}$ are mutually exclusive
statements; therefore, ``$\Lambda_{k_i}$ and $\Lambda_{k_j}$'' should
be false and, correspondingly, $\hat{\Lambda}_{k_i}
\hat{\Lambda}_{k_j} = 0$.  Equation (8) gives a logical meaning for
quantum operators representing physical observables. 

Let a physical system be in  state $\mid \psi >$, which may be an
eigenstate of some physical observable, $Q$, so $\mid \psi > \equiv
\mid q >$, where $q$ is a numerical value of $Q$ in this state.
According to (8), the average value of $K$ in state $\mid q >$, $< q
\mid \hat{K} \mid q >$, is equal to 
$$
< q \mid \hat{K} \mid q > \equiv tr \left ( \hat{\Lambda}_q \hat{K}
\right) = \sum_i k_i tr \left( \hat{\Lambda}_q \hat{\Lambda}_{k_i}
\right)~~, 
\eqno(9)
$$
\noindent where $tr (\hat{\Lambda}_q \hat{\Lambda}_{k_i} ) \equiv < q
\mid \hat{\Lambda}_{k_i} \mid  q > = \mid  < q \mid k_i > \mid^2$~~,
$$
\hat{\Lambda}_q =  \mid q >< q \mid ~,~~ \Lambda_q~:~~ {\mbox{``}}Q =
q{\mbox{''}}~. 
\eqno(10)
$$
\noindent $tr ( \hat{\Lambda}_q \hat{\Lambda}_{k_i})$ is the average
truth of $\Lambda_{k_i}$ when $\Lambda_q$ is true; but we can say as
well 
that it is the conditional truth of $\Lambda_{k_i}$, $Tr (\Lambda_q
\mid \Lambda_{k_i})$, under the
condition that $\Lambda_q$ is true; it would also be correct to say
that $tr ( \hat{\Lambda}_q \hat{\Lambda}_{k_i})$ is the level of
confidence in the value $K = k_i$ when the physical system is in state
$\mid q >$.  Thus,
$$
Tr ( \Lambda_q \mid \Lambda_{k_i} ) = tr \left( \hat{\Lambda}_q
\hat{\Lambda}_{k_i} \right)~~.
\eqno(11)
$$
\noindent (Note that for two elementary statements, $\Lambda_q$ and
$\Lambda_{k_i}$, $Tr ( \Lambda_q \mid \Lambda_{k_i} )= Tr (
\Lambda_{k_i} \mid \Lambda_q )$.)

The logical meaning of formula (9) is the following:  when calculating
the  average  $K$ in state $\mid q >$,
we need to take every $k_i$ with its level of confidence, and then
summarize the terms.

As a rule, the density matrices $\hat{\Lambda}_q$ and
$\hat{\Lambda}_{k_i}$ in (11) should be taken at the same time,
although there are cases in which it is not essential.  What is
essential, however, is that {\underline{both}} $\Lambda_q$ and
$\Lambda_{k_i}$ describe possible physical situations
{\underline{between}} measurements, i.e., when the physical system is not
observed.  If $\Lambda_q$ describes quantum state $\mid q >$ before the
measurement of observable $K$, which does not commute with $Q$,
and  $\Lambda_{k_i}$ describes a state created by a single
measurement, then it makes no sense to talk about the conditional
truth of $\Lambda_{k_i}$ in the state $\mid q >$ that has already been
destroyed  by the
measurement.  In this case $tr (\hat{\Lambda}_q \hat{\Lambda}_{k_i})$
equals  probability $w ( \Lambda_q \mid \Lambda_{k_i})$ that
$K = k_i$ will be the result of the measurement, if before this
measurement the system was in state $\mid q >$.  Although the formulae
are the same
$$
Tr ( \Lambda_q \mid \Lambda_{k_i} )  = w (\Lambda_q \mid
\Lambda_{k_i}) = tr (\hat{\Lambda}_q  \hat{\Lambda}_{k_i}) \equiv \mid
< q \mid k_i > \mid^2~~,
\eqno(12)
$$
\noindent the physical situations and hence the meanings are
different.  This difference is essential in explaining the main
peculiarity of the EPR-Bohm experiments:  the allegedly 
``faster-than-light'' mutual influence of two spatially separated
actions of measurements.

(Note that $Tr$ and $w$ in (12) are equal only because we chose truth
values 1 for ``true'' and 0 for ``false''.  If we followed the
choice given, say, in [4], equation (12) would be different.)

We see from this analysis that quantum density matrices of pure states
represent logical truths of statements describing  those states.  And
both density matrices and wave functions~---~the eigenvectors of
density matrices~---~contain information about conditional
truths of all possible logical statements about observables of a given
physical system.  If the system wave function is $\mid \psi
>$, then the conditional truth of a statement $M$ is equal to $Tr
(\Lambda_\psi \mid M) 
= tr ( \hat{\Lambda}_\psi \hat{M})$, where $\Lambda_\psi$ describes
the condition:  ``The system is in state $\mid \psi >$''.  Although
$tr (\hat{\Lambda}_\psi \hat{M})$ does not depend on the order of
cofactors, the order in $Tr (\Lambda_\psi \mid M)$ can be important: 
$\Lambda_\psi$ is always an elementary statement, $tr
\hat{\Lambda}_\psi = 1$, while $\hat{M}$ can be any complex
statement.  For any statement $M$, 
$$
0 \leq Tr ( \Lambda_\psi \mid M ) \leq 1~~.
\eqno(13)
$$
\noindent $M$ and $\Lambda_\psi$ need not be the same when $Tr = 1$,
since $M$ can be a disjunction containing $\Lambda_\psi$; they
are mutually exclusive when $Tr= 0$.

One type of complex statement plays an 
important role in the transition from quantum to
classical descriptions:
$$
\Delta \hat{I}_k = \sum^{i_2 = i_1 + N}_{i_1} \hat{\Lambda}_{k_i}~~,~~
N \gg 1~~,
\eqno(14)
$$
$$
\Delta I_k~:~~ {\mbox{``}}\Lambda_{k_1}~ {\mbox{or}}~ \Lambda_{k_2}~
{\mbox{or}}~ \ldots {\mbox{or}}~ \Lambda_{k_{1 + N}}{\mbox{''}}~~.
\eqno(15)
$$
\noindent (See Sec.~2 for proof that truth operator (14)  represents 
statement (15).)~~ (15) describes a
big interval of neighboring quantum numbers of observable $K$.
Consider $\Delta I_p$ 
and $\Delta I_q$, where $p$ and $q$ are generalized momentum and
coordinate.  Defining $p$ as a translational invariant
in $q$-space (see Sec.~3) we can find $\hat{\Lambda}_{p_i}$ and
$\hat{\Lambda}_{q_j}$ in the same representation.  It is then easy to see
that if intervals $\Delta q$, $\Delta p$ in (14) are such that
$\Delta S/\hbar \gg 1$, then $\Delta \hat{I}_p$ and $\Delta \hat{I}_q$
commute.  Obviously, classical trajectories are distinguished from
each other, if $\Delta S/\hbar \ll S/\hbar$, where $S$ is a typical
action of a given physical system.  Under these conditions, the
description becomes classical.  Thus, the classical picture of the world
appears when physical differences between micro-states inside 
macro-intervals (15) either are not essential or cannot be resolved,
and $1 \ll \Delta S/\hbar \ll S/\hbar$.

It follows from the above that quantum
Hilbert space is an information-space in the sense
explained earlier.

\bigskip\bigskip

\pagebreak

\noindent {\bf 2.~~Representations of complex statements. Logical
correlations between noncommuting observables.}

\bigskip

The simplest complex statement is a mixture of mutually  exclusive
statements.  If wave functions $\mid k_j
>, j = 1, 2, \ldots$, represent a full set of orthonormalized states,
then a mixture of  $\Lambda_{k_j}$'s is, by definition,  
a probabilistic distribution 
of statements $\Lambda_{k_j}, j = 1, 2, \ldots$;  $w_j$'s are
classical probabilities in the sense that they result from lack of
knowledge. 
We will denote the mixture of $N$ statements with probabilities $w_1,
\ldots, w_N$ by $\rho_{w_1, w_2, \ldots ,
w_N}$.  The corresponding truth operator is a mixed density matrix
(which can also be considered a probability  operator):
$$
\hat{\rho}_{w_1, w_2, \ldots , w_N} (x, x^\prime) = \sum_j w_j
\hat{\Lambda}_{k_j} (x_1 x^\prime)~~.
\eqno(16)
$$

Our next aim is to find the truth operators of complex statements
composed of more simple statements, by using  logical connectives;  
negation of a statement $M$ (denoted by $\bar{M}$); ``and''
(denoted by $\wedge$); ``or'' ($\vee$, inclusive); implication
($\rightarrow$); etc.  Quantifiers $\forall$, $\exists$ are not
discussed. 

The formulae for the truth operators of complex statements are well
known [2,3]; they coincide with the formulae for the probabilities of
complex events.  However, these formulae are not formally deduced in
[2,3]; moreover, it is assumed in [3] that {\underline{all}}
projectors can be interpreted as propositions. Thus far 
we have proved only that density matrices of pure states can
be interpreted as truth operators of corresponding propositions
(statements).  In this section, which relates to noncommuting
truth operators, it will be shown that when the necessary conditions
(24) are met, 
the formulae for complex statements are valid for noncommuting
observables. 

Let $M_1, M_2$ be two logical statements about quantum observables of a
given physical system.  We will first assume that their truth
operators commute, $[\hat{M}_1 , \hat{M}_2] \equiv \hat{M}_1 \hat{M}_2
- \hat{M}_2 \hat{M}_1 = 0$, and then find the conditions under which
the formulae for commuting truth operators are also valid for
noncommuting $\hat{M}_1$ and $\hat{M}_2$.

If $N$ mutually orthogonal states, $\mid \psi_n >$, $n= 1, 2, \ldots ,
N$, represent a basis of an $N$-dimensional complex vector space, then
there also exist $N$ mutually exclusive elementary statements,
$\Lambda_{\psi_n}, \hat{\Lambda}_{\psi_n} = \mid \psi_n > < \psi_n
\mid$, $tr \hat{\Lambda}_{\psi_n} = 1$.   They constitute a basic
language for constructing $2^N$ different statements, $M_i$, $i=1, 2,
\ldots , 2^N$, with the help of logical connectives.  All $M_i$'s commute,
$\hat{M}^2_i = \hat{M}_i$.  (There is a continuum of such statements
when $N = \infty$, but we will have no problems with it, if we do
not operate with the full set of these statements.)  $2^N$ is
the number of different distributions of $0$'s and $1$'s along the main
diagonals of density matrices representing different $M_i$'s in their
common diagonal representation.  The full set of these $2^N$
statements forms what we can call a classical logic of quantum
propositions in a given representation.  Then a transition to another
quantum representation (which is some unitary transformation in the
vector space) will provide us with another classical logic having 
its own $2^N$ mutually commuting different statements.  Most of the 
truth operators of these statements, however, will not commute with the truth operators of
the first representation; important exceptions are logical
tautologies, which are 
invariant in all representations.

Thus, in any quantum vector space with $N>1$, there is an infinite number of
mutually noncommuting classical logics that are transformed into each
other by symmetry transformations.  These quantum transformations
can be interpreted as redefinitions of logical truth values.  Assuming
$[\hat{M}_1, \hat{M}_2] = 0$  we consider only one of
such logics.

Any nonelementary truth-operator
containing $\hat{M}_1$ and $\hat{M}_2$, $\hat{M}^2_i = \hat{M}_i$, 
can be written as
$$
\hat{M} = a + b \hat{M}_1 + c \hat{M}_2 + d \hat{M}_1 \hat{M}_2~~.
\eqno(17)
$$
\noindent Indeed, all higher terms for commuting operators
obeying (2) can be reduced to (17).  Using (2), we get from (17) the
equation  $a^2 +b (b+2a) \hat{M}_1 + c
(c+2a) \hat{M}_2 +$ \\ $+ \left[ d^2 + 2 (ad + bc + bd + cd) \right]
\hat{M}_1 \hat{M}_2 = a + b \hat{M}_1 + c \hat{M}_2 + d \hat{M}_1
\hat{M}_2$. If $\hat{M}_{1,2} \neq \hat{0}, \hat{1}$ (these are zero
and identity operators), then
$$
a^2 = a,~ b^2 + 2 ab=b,~ c^2+2ac=c,~ d^2+2 (ad+bc+bd+cd)=d~~.
\eqno(18)
$$
\noindent Different solutions of these equations give us truth
operators for different logical connectives.  Identification of
these connectives is based on the conventional logical truth tables
given in textbooks.  We will demonstrate the process of identification
in cases of negation and conjunction. 

The symbol $\Rightarrow$ means here ``is represented
by the truth operator.''  For example:  $M \Rightarrow \hat{M}$ means
``$M$ is represented by the truth operator $\hat{M}$.''  

{\it Negation}.  $a=1, b=-1, c= d= 0$ (negation of $M_1$ denoted by
$\bar{M}_1$).
$$
M= \bar{M}_1 \Rightarrow \hat{1} - \hat{M}_1~~.
\eqno(19)
$$
\noindent  According to the truth table defining negation, if $M$
is true,  then $\bar{M}$ is false, and vice versa.  In quantum
mechanics, ``$M_1$ is true'' means that our physical system is in such
a state $\mid \psi >$ that $\hat{M}_1 \mid \psi > = \mid \psi >$.
Then $\hat{\bar{M}}_1 \mid \psi > = (1- \hat{M}_1 ) \mid \psi > = 0$,
which means that  formula (19) is correct.

When $N>2$ and $M_1 = \Lambda_{\varphi_n}$ (an elementary statement), 
$\bar{M}_1$ is not 
elementary since there is more than one state orthogonal to $\mid
\varphi >$.  Every $\Lambda_{\varphi_k}$, $k \neq n$, is a ``partial
negation'' of $\Lambda_{\varphi_n}$, since $\Lambda_{\varphi_k}$
declares that the system is in state $\mid \varphi_k >$, which
excludes the presence of state $\mid \varphi_n >$.  In (19),
${\hat{\bar{M}}}_1$ represents ``total negation'' (or, simply,
negation): if $M_1 = \Lambda_{\varphi_n}$, then
$$
\bar{M_1} = \mbox{``} \Lambda_{\varphi_1} \vee \Lambda_{\varphi_2}
\vee \ldots \vee \Lambda_{\varphi_{n-1}} \vee \Lambda_{\varphi_{n+1}}
\ldots \mbox{''}~~,~~~
\bar{M_1} \Rightarrow \hat{\bar{M}}_1 = \sum_{k \neq n}
\hat{\Lambda}_{\varphi_k}~~;
\eqno(20)
$$
\noindent see the explanation of ``disjunction'' below.

From (19) we get the conditional truth for negation when the physical
system is in state $\mid \psi >$:
$$
Tr (\Lambda_\psi \mid \bar{M}) = 1 - Tr ( \Lambda_\psi \mid M) = 1 -
tr ( \hat{\Lambda}_\psi \hat{M})~~.
\eqno(21)
$$
\noindent It is interesting to note that $M$ and $\bar{M}$ may have
the same conditional truth, 0.5.

{\it Conjunction}.  $a=b=c=0, d=1$.
$$
M=M_1 \wedge M_2 \Rightarrow \hat{M}_1 \hat{M}_2~;~~ Tr  \left(
\Lambda_\varphi \mid M_1 \wedge M_2 \right) = < \varphi \mid \hat{M}_1
\hat{M}_2 \mid \varphi > = tr \left( \hat{\Lambda}_\varphi \hat{M}_1
\hat{M}_2 \right)~~.
\eqno(22)
$$
\noindent $M$ is true in state $\mid \varphi >$ if both $M_1$ and
$M_2$ are true in that state, 
and false if at least one of them is false.  This corresponds to the
conventional truth table for ``and.''  In (22), $M_1$, $M_2$ can
be any statements 
about two coexisting quantum numbers, and $\mid \varphi >$ is an
arbitrary state; therefore, the commuting $\hat{M}_1, \hat{M}_2$ may
or may not
commute with $\hat{\Lambda}_\varphi$.

If $\hat{M}_1, \hat{M}_2$ do not commute, then
$\hat{M}_1 \hat{M}_2$ does not  represent any statement, since
$(\hat{M}_1 \hat{M}_2)^2 \neq \hat{M}_1 \hat{M}_2$.  The average,
$<\varphi \mid  \hat{M}_1 \hat{M}_2 \mid \varphi >$, where
$\hat{M}_1$, $\hat{M}_2$ are placed in a certain order, can of course 
exist.  And again, this average {\underline{can}} be interpreted as
the conditional truth of the statement ``$M_1$ and $M_2$'' in state $\mid
\varphi >$, if at least one of the truth operators, $\hat{M}_1$ or
$\hat{M}_2$,  commutes with $\Lambda_\varphi$.
If, say, $\hat{M}_1 \hat{\Lambda}_\varphi = \hat{\Lambda}_\varphi
\hat{M}_1$, then $\mid \varphi >$ is their common state and
$\hat{M}_1 \mid \varphi > = \lambda \mid \varphi >$,~~ $\lambda=0,1$.
Therefore, 
$$
Tr \left( \Lambda_\varphi \mid M_1 \wedge M_2 \right) = \lambda <
\varphi \mid \hat{M}_2 \mid \varphi>~;~~ \hat{M}_1 \mid \varphi > =
\lambda \mid \varphi >~,~~  \hat{M}_1 \hat{M}_2 \neq \hat{M}_2
\hat{M}_1~~.
\eqno(23)
$$
\noindent The case is clear:  when, for example, $\lambda =1$, and
$\hat{\Lambda}_\varphi \hat{M}_1 = \hat{M}_1 \hat{\Lambda}_\varphi$,
operator $\hat{M}_1$ does not influence state $\mid \varphi >$ and
therefore can be ignored.  When $\lambda = 0$, the statement ``$M_1
\wedge M_2$'' is false because $M_1$ is false in state $\mid \varphi
>$, in which  $\Lambda_\varphi$ is true; therefore, $Tr ( \Lambda_\varphi \mid M_1
\wedge M_2 ) = 0$.  In such cases the order of cofactors in
(23) can be arbitrary.  It is easy to see that the most general
conditions under 
which the order of $\hat{M}_1$ and $\hat{M}_2$ is not important
and, therefore, the statement ``$M_1$ and $M_2$'' {\it effectively}
makes sense, are described by the equations
$$
Tr \left( \hat{\Lambda}_\varphi \left[  \hat{M}_k \hat{M}_\ell \right]
\right) \equiv Tr \left( \left[ \hat{\Lambda}_\varphi \hat{M}_k
\right] \hat{M}_\ell \right) \equiv
Tr \left( \hat{M}_k \left[ \hat{M}_\ell \hat{\Lambda}_\varphi \right]
\right) = 0~~,~~~k \neq \ell = 1~ \mbox{or}~ 2~~.
\eqno(24)
$$

(24) provides the only logical restriction on 
meaningful logical correlations between any two noncommuting logical
observables, $\hat{M}_1$ and $\hat{M}_2$ (as well as between physical
observables $K_1,$~ $[\hat{K}_1 \hat{M}_1 ] = 0$, and $K_2,$~ $[\hat{K}_2
\hat{M}_2 ] = 0$), when the physical system is in state $\mid \psi >$.

Let state $\mid \psi >$ of a system of two identical particles consist of two
macroscopically separated branches.  So $\Lambda_\psi$ is ``nonlocal''
in the sense that it describes both branches simultaneously.  While
every particle may be localized somewhere (see the discussion about
this in Sec.~4), we cannot formulate {\underline{true}} statements
about their localization when the nonlocal $\Lambda_\psi$ is true.
But we can formulate hypothetical statements, $M_1$ and $M_2$, about
physically possible states in which $M_1$ and $M_2$ are localized in
different branches.  If $M_1$ and $M_2$ relate to relativistically
nonoverlapping places, $\hat{M}_1$ and $\hat{M}_2$ commute.
Therefore, the statement ``$M_1$ and $M_2$'' makes sense, and we can
calculate its conditional truth (22).  The value we will get belongs
to the nonobservable physical system.  (For more about this, see
Sec.~6.)

It can be proved,  however,  that if a 
system is in state $\mid \psi >$ before our observation, and the
result of  the 
observation is described by  statement $M$, then the probability of
such a result is equal to $tr ( \hat{\Lambda}_\psi \hat{M})$ [3], i.e.,
equal to the conditional truth of $M$ in state $\mid \psi >$.  Our
interpretation of this equality is 
that conditional truth is measurable and $\hat{M}$ is an
observable.  Now, if $M =$ ``$M_1$ and $M_2$'', then the measurement of $\hat{M}$
means the measurement of the correlation between $\hat{M}_1$ and
$\hat{M}_2$, which is a typical EPR situation.  Can the measured
correlation be  a result of a long-range, faster-than-light interaction
between two spatially separated measuring processes?  Of course not:  these
long-distance correlations were born together with state $\mid
\psi >$, existed before the measurement, and were only confirmed by
measurement.  (Nevertheless, the phenomenon of the faster-than-light
disappearance of the initial information contained in the pre-measured
state $\mid \psi >$ exists.  It will be explained in Sec.~4.)

In other complex statements investigated below we will meet the same 
restriction (24) on ``$M_1$ and $\hat{M}_2$'' when $\hat{M}_1$ and
$\hat{M}_2$ do not commute.

{\it Disjunction}.  $a=0, ~b=c=1, ~d= -1$.
$$
M = M_1 \vee M_2 \Rightarrow \hat{M}_1 + \hat{M}_2 - \hat{M}_1
\hat{M}_2~~.
\eqno(25)
$$
\noindent If $M_1$ and $M_2$ are mutually
exclusive, then $\hat{M}$ degenerates into the sum, $\hat{M} =
\hat{M}_1 + \hat{M}_2$; this explains formula (20).  Since the
disjunction of a full set of elementary statements
$\Lambda_{\psi_i}$, $\hat{\Lambda}_{\psi_i} = \mid \psi_i >< \psi_i
\mid$, $i = 1, 2, \ldots , N$, is merely the enumeration of all
mutually exclusive quantum states, such a disjunction is an invariant:
$\sum^N_{i=1} \hat{\Lambda}_{\psi_i} = \hat{1}$.  In the diagonal
representation, any $\hat{M}$ can be represented as some distribution
of 1's and 0's along the main diagonal of the matrix representing $M$.
The ``total
negation'' of $M$, $\bar{M} \Rightarrow \hat{\bar{M}}$, contains 0's
instead of 1's and 1's instead of 0's; so $\hat{\bar{M}} = \hat{1} -
\hat{M}$. 

The identity matrix, $\hat{1}$, is a purely logical invariant.  Other
invariants are defined by irreducible representations of physical
symmetries of a given physical system. For example, the statement
$\Lambda_{p_k}$, 
$$
\Lambda_{p_k}{\mbox{: ``The momentum of the physical system is equal
to $p_k$''}}~~,
$$
\noindent is a translational invariant.  The truth operator
$\hat{\Lambda}_{p_k} 
(q_1 q^\prime)$ in the coordinate $q$-representation is not diagonal,
that is, $\Lambda_{p_k}$ is not local.  Finding this matrix is one of
the methods  of quantization which will be demonstrated in the
next section.

{\it Exclusive ``or''}.  $a=0, b=c=1, d=-2$.
$$
M= M_1 ~\mbox{or}~ M_2 ~\mbox{but not both}~ \Rightarrow \hat{M}_1 +
\hat{M}_2 - 2 \hat{M}_1 \hat{M}_2~~.
\eqno(26)
$$

{\it Implication}.  $a=1, b=-1, c=0, d=1$.
$$
M= M_1 \rightarrow M_2  \Rightarrow \hat{1} - \hat{M}_1 + \hat{M}_1
\hat{M}_2~~.
\eqno(27)
$$
\noindent By definition,
$$
Tr (\Lambda_\varphi \mid M_1 \rightarrow M_2) {\mbox{~~and~~}} w \left(
\Lambda_\varphi \mid M_1 \rightarrow M_2 \right) = 1~, 
{\mbox{if}} \left\{ \begin{array}{c} \hat{M}_1 \mid \varphi > =
0~,~~\mbox{or} \\ \hat{M}_2 \mid \varphi > = \mid \varphi >~~.
\end{array} \right.
\eqno(28)
$$
\noindent In (28), the commutation of $\hat{M}_1$ and $\hat{M}_2$ is
not required, since (24) is fulfilled.  In general,
$$
w \left( \Lambda_\varphi \mid M_1 \rightarrow M_2 \right) = 1 - w
\left(
\Lambda_\varphi \mid M_1 \right) + w \left( \Lambda_\varphi \mid M_1
\wedge M_2 \right)~~. 
\eqno(29)
$$
\noindent So if, for example, $\hat{M}_1 \hat{\Lambda}_\varphi =
\hat{\Lambda}_\varphi \hat{M}_1$, then $w (\Lambda_\varphi \mid M_1
\rightarrow M_2 ) = 1 - \lambda + \lambda w (\Lambda_\varphi \mid
M_2)$, $\hat{M}_1 \mid \varphi > = \lambda \mid \varphi >$.  If $M_1$
is true in state $\mid \varphi >$, then $w (\Lambda_\varphi \mid M_1 
\rightarrow M_2) =<\varphi \mid \hat{M}_2
\mid \varphi >$, the same as for the conjunction $M_1 \wedge M_2$, and
for measuring $M_2$ in state $\mid \varphi >$.  Thus, if $M_1
\rightarrow M_2$ is true and $M_1$ is true, then $M_2$ is true (modus
ponens). 

{\it Equivalence}.  $ a=1, b=c=-1, d=2$.
$$
M = M_1 \longleftrightarrow M_2 \Rightarrow \hat{1} - \hat{M}_1 -
\hat{M}_2 + 2 \hat{M}_1 \hat{M}_2~~.
\eqno(30)
$$
\noindent If $M_1$ is true in state $\mid \psi >$, then, again, $w
(\Lambda_\psi \mid M_1 \leftrightarrow M_2) = < \psi \mid \hat{M}_2
\mid \psi >$.

Other choices of $a, b, c, d$ correspond  to the substitutions
$M_{1,2} \rightarrow M_{2,1}$, or $M_{1,2} \rightarrow \bar{M}_{1,2}$.

\bigskip\bigskip

\pagebreak

\noindent {\bf 3.~~Origins of quantization.  Appearance of
noncomputable functions.  Quantum Hilbert space.}

\bigskip

Here we will explain why the unification of a purely geometrical concept
of symmetry with a purely logical concept of truth is inconsistent with
the  classical concept of trajectories  and, together with the
assumption of linearity, leads to quantization.

Consider a pair of canonically conjugate observables,  generalized
momentum $p$ and coordinate $q$.  In
classical mechanics, there are two definitions of $p$:  (1) $p =
\stackrel{\bullet}{q} (m=1)$, and (2) $p$ is an integral of motion in
a homogeneous space.  These are two {\underline{dynamic}} definitions
of $p$; however, the second definition is geometrical as well, since
the homogeneity of space means its translational symmetry.  Let us now
discard the dynamic part of the second definition, and introduce a
purely geometrical concept of momentum:

\centerline{\it Momentum $p$ is an invariant of translational
symmetry.} 

Another conceptual departure from classical mechanics follows from a
realistic look at scientific theory.  We have nothing to say about
undescribed matter besides the fact that it exists; our scientific
theories are  about  our own 
logically organized descriptions of observed and nonobserved subjects,
and our predictions about results of future observations of the
latter.  Quantum mechanics is a partial realization of the above
geometrical and logical concepts. 

Let us introduce a logical statement: $\Lambda_{p_i}$ :
``$p = p_i$'', where $p_i$ is a possible numerical value of $p$, one
of the values which can appear in our observations.  We will analyze
the logical truth of $\Lambda_{p_i}$, $\hat{\Lambda}_{p_i}$, as a
basic variable.

Since $p_i$ is a translational invariant, the truth of the statement
``$p = p_i$'' is also a translational invariant.  Since we accept two
possible numerical values for truth, 1 and 0, we may express the truth
of $\Lambda_{p_i}$, $\hat{\Lambda}_{p_i}$, as  a matrix diagonal
operator, $\Lambda_{p_i} (p, p^\prime ) = \delta_{pp_i}
\delta_{p^\prime p_i}$, that should not depend on $q$.  
$\hat{\Lambda}_p$ is defined in $p$-space.  In classical mechanics we
have, however, another definition of $p$:  $p_i = dq/dt = \lim
(q_{k+1} - q_k)/(t_{k+1} - t_k)$, where $p$ is defined in $q$-space.
We should now reject this definition, assuming instead 
that, in $q$-space, the truth of $\Lambda_{p_i}$,
$\hat{\Lambda}_{p_i}$ must not depend on 
translations of coordinates, $q_k \rightarrow q_{k+1} = q_k + \delta
q$.  Calculation of matrix $\Lambda_{p_i} (q, q^\prime)$ under
this condition, plus two other conditions, $\hat{\Lambda}^2_{p_i} =
\hat{\Lambda}_{p_i}$ and $tr \hat{\Lambda}_{p_i} = 1$, gives:
$$
\Lambda_{p_i} (q, q^\prime) = A \exp \left\{ 2 \pi i (q^\prime -
q)/\lambda_i \right\}~~,
\eqno(31)
$$ 
\noindent where $\lambda_i$ is a wave length depending on $p_i$ and
$A$ is a normalization constant.  On the basis of an analogy between
$p$ and $q$ we conclude that $\lambda_i \propto
1/p_i$; and from experiment, $2 \pi/\lambda_i = p_i/\hbar$. Indeed,
$q$ is a translational invariant in $p$-space, like $p$ in $q$-space.
In classical mechanics, at a turning point, $\partial q/\partial p = 0$
along the trajectory in  phase space.  And in nonrelativistic
quantum mechanics, in order to measure a particle coordinate $q$, we
need to stop this particle for an instant, thus creating a turning
point and an instantaneous translational invariance along the
$p$-axis. 

Since $\hat{\Lambda}_{p_i} (p)$, with matrix elements
$\Lambda_{p_i} (p, p^\prime)$, and $\hat{\Lambda}_{p_i} (q)$,  with
matrix elements $\Lambda_{p_i} (q, q^\prime)$, both represent the truth
of statement ``$p = p_i$'' in different spaces, there should be a
connection between them.  The connection is:
$$
\Lambda_{p_i} (q,q^\prime) = \sum_{p, p^\prime} S (q,p) \Lambda_{p_i}
(p, p^\prime) S^+ (q^\prime , p^\prime )~~,
\eqno(32)
$$
$$
S(q,p) = \sqrt{A} \exp iqp/\hbar~;~~ S^+ = S^{-1}~~.
\eqno(33)
$$
\noindent Note that the unitarity (33) is derived from (31), and not
postulated. 

The eigenvectors of $\hat{\Lambda}_{p_i} (p)$ and $\hat{\Lambda}_{p_i}
(q)$, with the eigenvalue  $+1$, are $p$- and $q$-representations of
the same wave function, $\psi_{p_i}$, corresponding to  momentum $p =
p_i$:
$$
\psi_{p_k} (q) = S \psi_{p_k} (p) = \sqrt{A} \exp ip_kq/\hbar~;~~
\psi_{p_k} (p) = \delta_{pp_k}~~.
\eqno(34)
$$
\noindent The quantization is virtually completed.  Using, now, the
definition of $p$, 
$$
\hat{p} = \sum p_i \hat{\Lambda}_{p_i}~~,
\eqno(35)
$$
\noindent we will find that $\hat{p} = - i \hbar \partial/\partial q$.  We
have also got a complex, infinite (hence Hilbert) vector space with
unitary transformations in it.  As noted in the Introduction,
the very possibility of forming such spaces is contained in the
classical logic of propositions.  Indeed, if we represent truths of
propositions as matrices, then we can immediately conclude  that
geometrical 
transformations $\hat{M}^\prime = S \hat{M} S^{-1}$ give 
$(\hat{M}^\prime)^2 = \hat{M}^2~;~~ tr \hat{M}^\prime = tr \hat{M}$,
and $\hat{I}^\prime = \hat{I}$, where $\hat{I}$ represents a logical
tautology.  
Also, using the formulae for complex statements deduced in Sec.~2, and the
standard textbook tables of logical tautologies, 
we can confirm that every logical tautology is true in every state $\mid \psi
>$.  Or, using the technique of forming full sets of mutually
exclusive statements described in [3], we will see that all
tautologies can be represented as identity matrices.  

A crucial point in our deduction of quantization has been the separation of
a purely geometrical symmetry from particle dynamics, i.e., from 
symmetry and the very existence of a Hamiltonian.  Using as an example 
angular momentum, $(M_x , M_y , M_z)$ and rotational SU$_2$
symmetry, we will now show how noncomputable functions appear in such
mechanics.  These functions are  crucial to explaining 
the phenomenon of quantum indeterminism (Sec.~5).

$M_z$ is defined now only as an invariant of axial symmetry about the
$z$-axis; $M_x$ and $M_y$ are defined correspondingly.  Let $M_z = m$.
Then $m$ is an invariant of rotations around the $z$-axis, as is the
truth of the 
statement $\Lambda_m, \Lambda_m$: ``$M_z = m$''.  Using the same equations for
the truth of $\Lambda_m$, $\hat{\Lambda}_m$, that we used for
$\Lambda_{p_i}$ in the case of momentum $p = p_i$ and
applying the condition  of periodicity in this case, we get the
matrix $\Lambda_m (\varphi , \varphi^\prime)$, where $\varphi$ is the
angle of 
rotation around the $z$-axis,
$$
\Lambda_m (\varphi , \varphi^\prime) = \frac{1}{2 \pi} e^{im (\varphi^\prime -
\varphi)} = \mid m >< m \mid~~,
\eqno(36)
$$
$$
\mid m > = \frac{1}{\sqrt{2\pi}} e^{im\varphi}~~,~~~m~ \mbox{is integer.}
\eqno(37)
$$
\noindent Since angle $\varphi$ is uncertain (truth operator
$\hat{\Lambda}_m$ is not diagonal), the directions of the $x$- and
$y$-axes are also uncertain; therefore, $M_x$ and $M_y$ cannot
have exact numerical values simultaneously with $M_z$. 

Let us now measure $M_z/\hbar$, at first about one axis, $z_1$, and then
about another, $z_2$.  Let the maximal  number of
possible $M_z$-projections  equal  some $N$.  In SU$_2$, there
exists a rotation $G$ (around the axis perpendicular to both $z_1$ and
$z_2$) that transforms rotations $R_{z_1}$ into rotations  $R_{z_2}$, 
$$
R_{z_2} = GR_{z_1} G^{-1}~;~~ R_{z_2} R_{z_1} - R_{z_1} R_{z_2} \neq
0~~, 
\eqno(38)
$$
\noindent and $G$ can be represented in the form
$$
G = G (\alpha_{z_1 z_2})  = \exp (i \alpha_{z_1 z_2} \hat{g} )~~,
\eqno(39)
$$
\noindent $\hat{g}$ is a Hermitian operator.

The symmetry in this case means that the number and values of
the angular momentum projections onto axis $z_1$ are the same as the
number and values of projections onto the equivalent axis, $z_2$.
(This is radically different from classical mechanics.)  Suppose there
exists a computable function which maps one-to-one $(M_{z_1})_i
\rightarrow (M_{z_2})_j~,~~ i~,~~ j = 1, 2, \ldots , N$.  There are
$N$! versions of such mappings; every one
is a permutation, $i \rightarrow j; i, j = 1, 2, \ldots, N$.
Computability means that for every $\alpha_{z_1 z_2}$ we can calculate
the corresponding permutation.

Our proof of noncomputability will be rather informal, and will not
address the theory of recursive functions or Turing machines.  
Suppose that at least for some $\alpha_{z_1 z_2}$ the permutation is
not the identity permutation, i.e., not $i \rightarrow i, i = 1,
\ldots , N$.  Let us divide $\alpha_{z_1 z_2}$ into $N$! intervals,
$\Delta \alpha = \alpha_{z_1 z_2}/N$!.  So,
$$
G(\alpha_{z_1 z_2}) = \left( G (\Delta \alpha )\right)^{N!}~~.
$$ 
\noindent Every small rotation $G(\Delta \alpha)$ corresponds to some
permutation (depending only on $\Delta \alpha$), $P(\Delta \alpha)$,
and the full rotation to the full permutation, i.e.,
$$
P^{N!} (\Delta \alpha) = P (\alpha_{z_1 z_2})~~.
$$
\noindent However, $P^{N!} (\Delta \alpha) = \hat{1}$, identity 
transformation.  {\it Contradiction.}

Thus, there does not exist a one-to-one computable function that
translates certain $M_{z_1}$ projections into certain $M_{z_2}$
projections.  (The simplest case $M_z =
\pm \frac{1}{2}$ is analyzed in Sec.~5.)

It does not follow immediately from this result that quantum mechanics
is indeterministic.  We need to show why and how, when measuring the
$z_2$-projection of angular momentum initially directed along the 
$z_1$-axis, we should and can exclude the possibility of quantum
superpositions of different $z_2$-projections.  We need to explain the
difference between an apparatus as a measuring device, and as a
physical target.  This will be done in Sec.~5.

\bigskip\bigskip

\pagebreak

\noindent {\bf 4.~~Stern-Gerlach experiment.  Quantum nonlocality
(uncertainty). Measurement of nonlocal observables.}

\bigskip

In the Stern-Gerlach experiment (SG), two branches of the 
final state of a heavy atom $A$ having 1/2-spin $S$ are spatially
separated.  This is a necessary condition for measuring the
probability of the $S_Z$-projection.  Axis $\vec{Z}$ in Fig.~1 is
perpendicular to the central trajectory.  Atoms are polarized along
$\vec{Z}_1$ or $\vec{Z}_2$.  In Fig.~2 we can see an original result
of the beam splitting [5].

The initial wave packet before entering the magnet is
$$
\psi_0 = u^0 \left( \vec{R} t \right) \sum_m c_m \phi_m \left( \vec{r}
\right)~,~~ m= \pm 1/2~~,
\eqno(40)
$$
\noindent and after exiting from it,
$$
\psi = \sum_m c_m u_m \left( \vec{R} t \right) \phi_m \left( \vec{r}
\right)~~,
\eqno(41)
$$
\noindent where $\phi_m (\vec{r})$ describes the internal atom motion,
and $u_m (\vec{R}t)$ the center-of-mass motion; $u_{1/2}$
and $u_{-1/2}$ do not overlap spatially:
$$
u_m \left( \vec{R} t \right) u^\ast_{m^\prime} \left( \vec{R} t
\right) = u_m \left( \vec{R}^\prime t \right) u^\ast_{m^\prime} \left(
\vec{R}^\prime t \right) = 0~,~~ m \neq m^\prime~~.
\eqno(42)
$$

As pointed out in [6], if we measure only local observables, then
density matrix $\hat{\rho}$, corresponding
to (41), can be effectively replaced by classical density matrix 
$\rho_c$, in which off-diagonal terms of $\hat{\rho}$ are omitted.
This can explain the collapse of quantum states in most
cases: during  measurement of local observables, information about
phases is lost.  Before discussing measurement of nonlocal observables 
like $\hat{\rho}$, namely of their off-diagonal terms, we will first
discuss the definition of quantum nonlocality. 

We will call a set of eigenvalues of an observable $K$, $\{ k_1 ,
k_2 , k_3 , \ldots \}$, a $K$-space.  We will call $K$ {\it local in a
given $K$-space}, or simply local, if $K$ is equal to one of its
eigenvalues from this space, $K=k_j$, $k_j \epsilon \{ k_1, k_2 ,
\ldots \}$.  If $K$ is local in a $K$-space, $Q$ is local in a
$Q$-space, and there exists a one-to-one correspondence, $q_j
\leftrightarrow k_i$ for every $q_j$ and every $k_j$, then we will say
that both $K$ and $Q$ are nonlocal in both $K$- and $Q$-spaces.
For example, in SG, in state (41), $S_Z$ is local in $\vec{R}$-space.  
Obviously, $K$ is local if it belongs to a physical system whose
quantum state is an eigenvector of $K$, $\mid k_j >$, $k_j \epsilon \{
k_1 , \ldots \}$.  The question is whether $K$ is local in 
superposition $\mid \psi > = \sum a_j \mid k_j >$, $a_j \neq 0, 1$.

We may suggest that despite the quantum uncertainty 
$\Delta K$ in such a state, $K$ {\underline{has}} a
certain (although unknown) numerical value, and therefore that the
physical  system,
though in a state of a superposition, nevertheless occupies  some
(unknown to us) point in $K$-space.  
However, there is no way either to prove or disprove such a suggestion.  Let $\mid \psi
>$ be an eigenstate of an observable $P$ not commuting with
$K$, $\hat{P} \mid \psi > = p_i \mid \psi >$, $[\hat{P}, \hat{K} ]
\neq 0$.  As shown in the previous section, there does not
exist a computable function mapping $P$-space one-to-one onto
$K$-space.  Noncomputable functions (see [7], for example) cannot be
described algorithmically.  We have a typical quantum effect: quantum
nonlocality.  This nonlocality  does not mean that a physical system is
``smeared'' in a given space; it means only that there are no
algorithms to prove or disprove the statement, ``The system is located
at a certain point.''   On one hand we know, after G\"odel's work
[8], (see also [4], [7]), that such a statement in principle can  be
true; on 
the other hand, the question, ``Where is the system located?'' makes
no sense.

Nevertheless, it is reasonable to call $K$ 
{\it nonlocal in $K$-space} in state $\mid \psi >$, if $\hat{K}
\hat{P} - \hat{P} \hat{K} \neq 0$ and $P \mid \psi > = p_j \mid \psi
>$.  Thus, if $K$ is nonlocal, the quantum state is the superposition 
$\sum a_j \mid k_j >$, $a_j \neq 0, 1$, and vice versa.

But quantum nonlocality so defined is merely the familiar quantum
uncertainty, $\Delta K$, which depends on the state of the system.
This clarification of the term will help us understand how to
measure off-diagonal terms of density matrix $\hat{\rho}$.  We need to
clarify also the concept of measurement.  
There are two radically different cases of the interaction of a
physical system with a detector designed to measure $K$.  

\begin{enumerate}

\item The result of the interaction can be formulated as ``$K  =
k_j$'', where $k_j$ is an exact numerical value.  In such a case the
detector is a part of a measuring apparatus; but this requires that
there be a special logical procedure, used by the
researcher, that permits him/her to distinguish different numerical
values of $K$, $k_i$, $k_j$, $k_j \neq k_i$.  In this case $K$ is
local  (as are the truths of all statements $\Lambda_{k_j}$, $j = 1,
2, \ldots$, $\hat{\Lambda}_{k_j} = \mid k_j >< k_j \mid$), for all
states $\mid k_j >$, $i = 1, 2, \ldots$, created by the measuring
apparatus.  

And if a pre-measured  state $\mid \psi >$ is an eigenstate of $Q$,
and 
$[\hat{Q} \hat{K} ] \neq 0$, as is usual in measurements, then
neither $K$ nor any $\Lambda_{k_j}$ is local in the $K$-space of the
pre-measured system.  Let $\hat{Q} \mid \psi >= q_i \mid \psi >$.  As
was explained, there is no computable function for one-to-one
mapping of  $Q$-space onto $K$-space; therefore, there are no logical
connections between statements $\Lambda_\psi \equiv \Lambda_{q_i}$ 
and $\Lambda_{k_j}$.  This means that transition $\hat{\Lambda}_\psi
\rightarrow \hat{\Lambda}_{k_j}$, caused by a single measurement, is not
deterministic.  The probability is equal to $tr ( \hat{\Lambda}_\psi
\hat{\Lambda}_{k_j})$.  If (and only if) $\hat{\Lambda}_\psi$ is
expressed in the $K$-representation, where $\hat{\Lambda}_{k_j}$ is
diagonal, then the off-diagonal terms of $\hat{\Lambda}_\psi$,
$(\Lambda_\psi)_{ik} = a_i a^\ast_k$, $i \neq k$, $\mid \psi > = \sum
a_j \mid k_j >$ play no role and can
be omitted.  So in our predictions of the results of this measurement,
$\hat{\rho}_\psi \equiv \hat{\Lambda}_\psi$ can indeed be replaced
by the classical $\rho_c : (\hat{\rho})_{ik}
\rightarrow (\rho_c)_{ik} = \delta_{ik} a_i a^\ast_k$.  
However, we will see that in some important
cases we should express $\hat{\rho}_\psi$ in a representation
different from the $K$-representation; in such cases the off-diagonal
terms of $\hat{\rho}_\psi$ cannot be omitted.

\indent (The logical procedure which permits us to distinguish between
$k_i$ and $k_j$, $i \neq j$, emerging from the
measurement, and therefore forbids interferences, will be discussed
in the next section.)

\item If such a logical procedure is not used, then the detector is
not a part of the measuring  apparatus, but is a target.  Despite the
fact that the apparatus is constructed to measure  observable $K$,
the quantum state emerging after the interaction of the
system with the detector is not in general an eigenstate of $K$,
i.e., $K$ is not local either in the initial or in the final quantum
states. 

\end{enumerate}

We now represent two methods of measuring nonlocal observables,
i.e., their off-diagonal terms, using the SG example.

In the first method, to measure off-diagonal terms of
$\hat{\rho}$ we need simply compare the truths of two statements,
$\Lambda_1$ and $\Lambda_2$,
$$
\Lambda_1 : \mbox{``} S_{Z_1} = \frac{1}{2} \mbox{''} \Rightarrow
\hat{\rho}^{(1)}~;~~ \Lambda_2 : \mbox{``} S_{Z_2} = \frac{1}{2}
\mbox{''} \Rightarrow \hat{\rho}^{(2)}~~,
\eqno(44)
$$
\noindent $\vec{Z}_1$ and $\vec{Z}_2$ are shown in Fig.~1.  The
probability that $\Lambda_2$ is true when $\Lambda_1$ is true (or vice
versa),
$$
\begin{array}{c}
w \left( \Lambda_1 \mid \Lambda_2 \right) = tr  \hat{\rho}^{(1)}
\hat{\rho}^{(2)} = \int d^3 r d^3 r^\prime d^3 R d^3 R^\prime \times
\\
\sum_{mm^\prime} c^{(1)}_m c^{\ast (1)}_{m^\prime} u_m \left(
\vec{R} t
\right) \phi_m \left( \vec{r} \right) u^\ast_{m^\prime} \left(
\vec{R}^\prime
t \right) \phi^\ast_{m^\prime} \left( \vec{r}^\prime \right)
\sum_{nn^\prime} c^{(2)}_n c^{\ast (2)}_{n^\prime} u_n \left(
\vec{R}^\prime t \right) \phi_n \left( \vec{r}^\prime \right)
u^\ast_{n^\prime} \left( \vec{R} t \right) \phi^\ast_{n^\prime} \left(
\vec{r} \right) = \\
= \mid c^{(1)}_{1/2} \mid^2 \mid c^{(2)}_{1/2} \mid^2 + \mid
c^{(1)}_{-1/2} \mid^2 \mid c^{(2)}_{-1/2} \mid^2 + \mbox{~
Interference term}~~,
\end{array}
\eqno(45)
$$
$$
\mbox{Interference term}~ = c^{(1)}_{1/2} c^{\ast (1)}_{-1/2} c^{\ast
(2)}_{1/2} c^{(2)}_{- 1/2} + c. c.
\eqno(46)
$$

In (45), (46) $c^{(k)}_{\pm 1/2}$ denotes the amplitude of the $S_Z =
\pm 1/2$ component of wave function $\psi^{(k)}$ that corresponds to
the initial polarization of spin along $\vec{Z}_k$.  In Fig.~1
$(\vec{Z}_k \vec{Z}) = \cos \theta_k$.  In SG, $c^{(k)}_{1/2} = e^{i
\varphi_k/2} \cdot \cos
\theta_k/2$, $c^{(k)}_{1/2} = \sin \theta_k/2$.  The interference term
is equal to $\frac{1}{2} \sin \theta_1 \sin \theta_2 \cdot \cos
\frac{\varphi_2 - \varphi_1}{2}$.  We can measure it by
comparing different probabilities $w (\theta_k) = \mid c^{(k)}_{1/2}
\mid^2$ in experiments with three different initial polarizations:
$\theta_1$, $\theta_2$, $\theta_3 = (\theta_2 - \theta_1)$, see
Fig.~2.  Then
$$
\mbox{Interference term}~ = w \left( \theta_2 - \theta_1 \right) - w
\left( \theta_1 \right) w \left( \theta_2 \right) - \left[ 1-w \left(
\theta_1 \right) \right] \left[ 1-w \left( \theta_2 \right) \right]~~.
\eqno(47)
$$
\noindent In these three measurements, we could keep the initial polarization
fixed along the $\vec{Z}_1$-axis, and rotate only the SG-magnet, making three
different angles between the $\vec{Z}_1$-axis and $\vec{Z}$-axis of the
magnet; let the three directions of the $Z$-axis be $\vec{Z}^{(1)}
(\theta^{(1)} = \theta_1 )$; 
$\vec{Z}^{(2)} (\theta^{(2)} = \theta_2 )$; $\vec{Z}^{(3)}
(\theta^{(3)} = \theta_2 - \theta_1 )$; $\cos \theta^{(k)} = (
\vec{Z}_1 \vec{Z}^{(k)})$.  We always measure the probability of the
$S_Z = \frac{1}{2}$ projection.  So we have three
$\hat{S}_Z$-operators, $\hat{S}^{(1)}_Z$, $\hat{S}^{(2)}_Z$, and
$\hat{S}^{(3)}_Z$, which do not commute.  Therefore, each $S_Z^{(k)}$
is local in states emerging after its own measurement, but nonlocal
in states emerging after the measurement of other $S_Z$'s.  The
initial pre-measurement state (which is always the same) collapses into
three types of final, post-measurement states, which are described by three
types of statements: $\Lambda^{(1)}_{\frac{1}{2}}~,~~
\Lambda^{(1)}_{ - \frac{1}{2}}~;~~ \Lambda^{(2)}_{\frac{1}{2}}~,~~
\Lambda^{(2)}_{ - \frac{1}{2}}~;~~ \Lambda^{(3)}_{\frac{1}{2}}~,~~
\Lambda^{(3)}_{ - \frac{1}{2}}$; their truth operators do not commute.
$$
\hat{\Lambda}^{(k)}_m \hat{\Lambda}^{(k^\prime)}_{m^\prime} -
\hat{\Lambda}^{(k^\prime)}_{m^\prime} \hat{\Lambda}^{(k)}_m = \left\{
\begin{array}{c} 0~,~~ k = k^\prime  \\ {\rm{something,}}~ k \neq
k^\prime  \end{array} \right. ~~. 
\eqno(48)
$$

The initial density matrix, $\mid \psi ><\psi \mid$, collapses all
three times but into different final density matrices
$\hat{\Lambda}^{(k)}_m$, $k = 1, 2, 3, m = \pm \frac{1}{2}$.  Each
time, the off-diagonal terms play
no role, but being expressed in the same
representation, they are partially restored in (47).

In the second method the final space of numerical values
emerging from the measurement is not changed, but it is neither the 
$S_Z$-space of the first SG-magnet (see Fig.~3), nor the 
$S_{Z_2}$-space of the second SG-magnet; it is the space resulting from
the interference of $\rho^{(1)}$ and $\rho^{(2)}$, see below.  None of
the $S_Z$'s corresponding to the two magnets are local in the final
states.  

In Fig.~4 the atom polarized along the $\vec{Z}_1$-axis is moving
toward the reader.  After passing the first SG magnet with symmetry
axis $\vec{Z}$, the wave function is split into upper and lower
branches along the $\vec{Z}$-axis with amplitudes $a$ and $b$, so the
density matrix~---~the  truth operator of logical statement about the
atom after the first SG,
$$
\hat{\rho}^{(1)} = \left( \begin{array}{cc} \mid a \mid^2 & ab^\ast \\
a^\ast b & \mid b \mid^2 \end{array} \right)~~.
\eqno(49)
$$
\noindent After passing the second magnet, both $a$- and $b$-branches
are split again into two branches, along the $\vec{Z}_2$-axis, with
the final amplitudes shown in Fig.~4, where the density matrix
$$
\hat{\rho}^{(2)} = \left( \begin{array}{cc} \mid c \mid^2 & cd^\ast \\
c^\ast d & \mid d \mid^2 \end{array} \right)~, \hat{\rho}^{(1)}
\hat{\rho}^{(2)} - \hat{\rho}^{(2)} \hat{\rho}^{(1)} \neq 0~~,
\eqno(50)
$$
\noindent and, by definition, corresponds to the transition from
$\vec{Z}_2$ to $\vec{Z}$. 

If, now, detectors $D_A$, $D_B$ register superpositions 
of the upper and lower branches inside the circles of Fig.~4, then we
observe 
$$
\mid A \mid^2 = tr \left( \hat{\rho}^{(1)} \hat{\rho}^{(2)} \right)~~.
\eqno(51)
$$

If $a= \stackrel{i \psi_1/2}{e \cdot \cos} \theta_1/2~,~~ c =
\stackrel{i \psi_2/2}{e \cdot \cos} \theta_2/2~$, then
$$
\mid A \mid^2 = \cos^2 \frac{\theta_1}{2} \cos^2 \frac{\theta_2}{2} +
\sin^2 \frac{\theta_2}{2} \sin^2 \frac{\theta_2}{2} +
{\rm{~~Interference ~term}}~~.
\eqno(52)
$$
\noindent Here we have interference between upper and lower
spots:  
$$
{\rm {Interference ~term}} = \frac{1}{2} \sin \theta_1 \sin \theta_2
\cos \frac{\psi_2 - \psi_1}{2}~~.
\eqno(53)
$$

We see that the interference of probabilities appears in (45), (47),
and (52) if (and only if) the experiment is constructed such that neither
$\hat{\rho}^{(1)}$ nor $\hat{\rho}^{(2)}$ in $tr (\hat{\rho}^{(1)}
\hat{\rho}^{(2)} )$ is expressed in its own diagonal representation.

\bigskip\bigskip

\pagebreak

\noindent {\bf 5.~~Measurement.  Indeterminism.  Collapse of wave
functions.  Quantum $\rightarrow$ classical transition.}

\bigskip

The Schr\"odinger equation describes a smooth deterministic development
of wave functions and density matrices.  Indeterminism and sudden
collapse of wave functions are observed only in measurements.  Why
does this happen?  What is the difference between a detector as a
physical target and a detector as  part of  a measuring apparatus?

A logical system processing the result of a measurement must, inter
alia, use a macroscopic scale whose ordered marks satisfy the
following conditions:
\begin{enumerate}
\item[(a)] They are separable;
\item[(b)] They are equal, in the sense that they do not provide any
service but marking;
\item[(c)] They have no inner structure; and
\item[(d)] There exists a one-to-one correspondence between them and
numerical  values of a measured
observable when the representation of the measured
states corresponds to the choice of apparatus.

\end{enumerate}

Exactly to satisfy conditions (a) and (d),  branches $S_Z =
\frac{1}{2}$ and $S_Z = - \frac{1}{2}$ of the SG experiment are
macroscopically separated in space.

The macroscopic separation of two wave packets in an SG
magnet becomes possible only because the following quasiclassical
condition is met:
$$
\mid \frac{d \lambda_z}{dz}  \mid = \frac{\hbar m_A}{F^2T^3} \ll 1~,~~
{\rm {or}}~~ \frac{S}{\hbar} \sim \frac{zp_z}{\hbar} \sim
\frac{F^2T^3}{\hbar m_A} \gg 1~~,
\eqno(54)
$$
\noindent where $\lambda_z$ is the de Broglie wave length, $\lambda_z
= \hbar/p_z$, $F = \mid \mu_e \frac{\partial B_Z}{\partial Z} \mid$,
$\mu_e$ is the electron magnetic moment, $B_Z$ is the $Z$-component of
the magnetic field; $m_A$ is the mass of the atom; and $T$ is 
flight time  through the magnet.  Condition (54) permits us not
only to separate two branches of the same wave function, but also to
observe macroscopic intervals (spots) (see Fig.~2) in both branches,
$\Delta z$, $\Delta p_z$, such that 
$$
1 \ll \frac{\Delta S}{\hbar} \ll \frac{S}{\hbar}~~,
\eqno(55)
$$
\noindent without distinguishing among different wave functions of
different atoms inside $\Delta S$.  The logical system of the
measuring procedure does not distinguish among them, in order to
satisfy conditions (b) and (c).

Consider an interaction between an atom and detectors in the SG
experiment, without taking into account detector efficiency.  After an
interaction occurs, 
but before it is registered by the logical system of the measurement
procedure, a common (atom  + detector)
wave function is a superposition,
$$
\psi_{com} = \sum_{m= \pm \frac{1}{2}} c_m \sum^N_k \alpha_{mk}
\psi^a_{mk} \phi^d_{mk}~~,
\eqno(56)
$$
\noindent where $c_m$ is taken from (41), $\sum_k \mid \alpha_{mk}
\mid^2 = 1$, $\psi^a_{mk}$ is one of many possible upper $(m =
\frac{1}{2})$ or lower $(m = - \frac{1}{2})$ atom eigenstates after
an interaction with a detector, and $\phi^d_{mk}$ is one of the detector
eigenstates; $N \gg 1$.  A change in the state of the
detectors is observed during measurement.  Without such an
observation, (56) is a pure 
quantum state.  Leaning on the examples given in the previous section
we can claim that, at least in principle, some experiments can give us
information about phases of $c_m \alpha_{mk}$, even if there are
many unknown phases.  
This information is irreversibly lost only when we directly count 
atom-detector interactions.

When events are counted, the logical system  of the measurement
procedure 
deliberately does not distinguish among different  events inside the
same (upper or lower) spots, $\Delta S_{\pm \frac{1}{2}}$, which obey
(55); $S$ is the action related to the event.  This means that
instead of considering precise statements about final states,
$\Lambda_{mk}$, with their truth operators,
$$
\hat{\Lambda}_{mk} = \mid \psi^a_{mk} \phi^d_{mk} >< \psi^a_{mk}
\phi^d_{mk} \mid~,~~ m = \pm \frac{1}{2}~,~~ k = 1, 2, \ldots~~, 
\eqno(57)  
$$
\noindent the logical system considers only complex statements, which
we can call in the SG case ``Spin up'' and ``Spin down'':
$$
{\rm{Spin}}^{\rm{up}}_{\rm{down}}:~~ {\rm{``Either}}~~ \Lambda_{\pm
\frac{1}{2}, 1}~,~~{\rm{or}}~ 
\Lambda_{\pm \frac{1}{2}, 2}~,~~{\rm{or}}~\Lambda_{\pm \frac{1}{2}, 3}~,
\ldots {\mbox{~''}}
\eqno(58)
$$

These statements are represented by truth operators (see formula (25) for
the case of 
mutually exclusive statements):
$$
{\rm{Spin}}^{\rm{up}}_{\rm{down}} \Rightarrow  \sum^N_k \mid
\psi^a_{\pm \frac{1}{2}k}
\phi^d_{\frac{1}{2}k} >< \psi^a_{\pm \frac{1}{2}k} \phi^d_{\frac{1}{2}k}
\mid \equiv \sum^N_k \hat{\Lambda}_{\pm \frac{1}{2}k}~~,
\eqno(59)
$$

If we now calculate the eigenvectors of these ``spin up'' and ``spin
down'' truth operators,  $V^{up}$ and
$V_{down}$,  we realize that these eigenvectors
{\underline{cannot interfere}}.
$$
{\rm{Spin}}^{\rm{up}}_{\rm{down}} = \sum_k  A_{\pm \frac{1}{2} k} \mid
\psi^a_{\pm \frac{1}{2}k} \phi^d_{\pm \frac{1}{2}k} >~~,
\eqno(60)
$$
\noindent where $A_{\frac{1}{2} k}$, $A_{-\frac{1}{2} k}$ are
completely arbitrary, so their phases are not defined.

Thus, conditions (a), (b), (c), and (d) are satisfied.  The transition from a
``micro'' to a ``macro'' description, essential for measurement,
includes fulfillment of physical condition (55); choice of an
apparatus; a decision  to ignore
details inside $\Delta S$ (or not to have physical tools resolve
them); assigning distinguishing names to  
different $\Delta S$-spots (like those in Fig.~2) separated in some
not always coordinate space; and a logical system of counting events. 

The unpredictability of results of measurement can be seen now
even from the comparison of dimensions.  We have  only two possible
{\underline{measured}} values, $S_Z = \frac{1}{2}$, $S_Z = -
\frac{1}{2}$, with no superpositions, while the set of possible
{\underline{initial}} states 
contains in addition two phases which appear in quantum mechanics in
return for the lost classical projections $S_X$ and $S_Y$.  The phase
values belong to the continuum, and there is no way to get a
deterministic mapping of all possible initial states onto two states,
$S_Z = \pm \frac{1}{2}$.  (A simple, more formal proof is given below.)

We know, after all, that some mapping occurs since we do get 
results of our measurements, though unpredictable
ones.  The  explanation is 
that the functions performing such mapping are noncomputable.

Suppose the opposite, namely, that there exists a computable
one-to-one function
in the SG case, i.e., a function mapping $S_{Z(\theta)}$ onto $S_Z$, 
where $S_Z$ is represented by the pointer positions: 
$$ 
S_Z = f \left(\theta~,~~S_{Z(\theta)} \right)~~.
\eqno(61)
$$
\noindent Here we use $Z(\theta)$ instead of the $\vec{Z}_K$ of Fig.~1,
$\vec{Z}_K \equiv Z (\theta_K)$, $Z \equiv Z(0)$.  $S_{Z(\theta)}$ is
the spin projection along axis $Z(\theta)$.  Only two
(invariant) eigenvalues are permitted, because $N=2$, 
$S_{Z(\theta)} = \pm
1/2$; therefore, there are only two possible rules for every $\theta$: 
$$
f \left( \theta~,~~S_{Z (\theta)} \right) = \pm S_{Z (\theta)}~~,
\eqno(62)
$$
\noindent and we need simply determine the correct sign.  It is
obvious (and this is our postulate) that for $\theta = \pi$,
$$
f \left( \pi~,~ S_{Z (\pi)} \right) = -S_{Z (\pi)}~~.
\eqno(63)
$$
\noindent Another sign would have made no sense.   Let $\theta_1 =
\pi/2$ and $\theta_2 - \theta_1 = \pi/2$ 
correspond to two consecutive mappings.  Rotation symmetry (about the
axis perpendicular to Fig.~1) demands that only relative
directions of axes are important, not their absolute directions
in space.  Thus, function $f$ can depend only on the angle between
axes.  Taking into account that the initial projection 
$S^{in}_{Z (\frac{\pi}{2})}$ relative to axis $Z
(\frac{\pi}{2})$
is equal to $S_{Z (\pi)}$ relative to $Z$, this gives:
$$
S_{Z \left( \pi/2 \right) } = f \left( \frac{\pi}{2}~,~~ S_{Z \left(
\pi \right)} \right) = \pm S_{Z \left( \pi \right)}~~,
\eqno(64)
$$
$$
S_{Z(0)} = f \left( \frac{\pi}{2}~,~~ S_{Z \left( \frac{\pi}{2}
\right)} \right) = \pm S_{Z \left( \pi/2 \right)} = + S_{Z \left( \pi
\right)}~~.
\eqno(65)
$$
\noindent But (65) contradicts (63).

Therefore, there does not exist a computable function that can perform 
the one-to-one mapping.  We can say also that there is no computable
one-to-one 
mutual translation of statements describing different directions of a
spin (with two obvious exceptions, $\theta = 0, \pi$).

Three things were crucial for the above result: 
mapping being permitted only onto $S_Z = \pm 1/2$ and not onto  
superpositions; the existence of rotation symmetry; and the invariance
of eigenvalues, that is, a property of symmetry in a complex vector
space. 

We see that the phenomenon of noncomputability relates only to
measurements, 
and leads to two effects:  (1) unpredictability of results of
measurements, and (2) collapse of measured states.

The effect of unpredictability (indeterminisim) of measurements
gives the basis for the statistical interpretation of wave functions.
The formula for the probability of a result $K= k_j$, when an initial
state is $\mid \psi >$, $w(\Lambda_\psi \mid \Lambda_{k_j} ) = tr  (
\hat{\Lambda}_\psi \hat{\Lambda}_{k_j}) \equiv  \mid < \psi \mid k_j >
\mid^2$, can be derived using the basic properties of probabilities, 
plus the assumption that $w$ should depend on both
$\hat{\Lambda}_\psi$ and $\hat{\Lambda}_{k_j}$, and only on them [3].
Since the properties of truth values are similar to the properties of
probabilities (for example, if both $\Lambda_\psi$ and $\Lambda_{k_j}$
are true in some state $\mid \varphi >$, then ``$\Lambda_\psi$ and
$\Lambda_{k_j}$'' is true, i.e., $\hat{\Lambda}_\psi
\hat{\Lambda}_{k_j} = 1$; and if $\Lambda_\psi$ and $\Lambda_{k_j}$
are mutually exclusive, then $\hat{\Lambda}_\psi \hat{\Lambda}_{k_j} =
0$), the formula for the truth of ``$\Lambda_\psi$ and
$\Lambda_{k_j}$'' coincides with 
the formula for the corresponding probability.  However, the concept
of probability is not valid for nonobserved systems, and the concept
of truth is purely logical and does not depend on measuring procedures. 

As for the collapse of a state $\mid \psi >$, $\mid \psi > \rightarrow
\mid k_j >$, the question arises of when and where this collapse
occurs.  
The answer is the following.  Since the collapse  is merely the 
creation of an unpredictable new description of a measured system, it
can occur only when and where the
logical system processing a given individual  measurement formulates
the result  ``$K = k_j$.''  Only after that can the  information (in
Shannon's sense) about this result be transmitted to other places.

\bigskip\bigskip

\pagebreak

\noindent {\bf 6.~~EPR-Bohm Experiment.  The main questions and
answers.} 

\bigskip

The principal setup of one of the EPR-Bohm experiments is shown in
Fig.~5.  (In real experiments, Stern-Gerlach magnets have never been
used as analyzers.  The principal setup used here for the sake of
simplicity  is taken from the original
paper [9].)  The orientations of the two analyzers, $\vec{Z}_a$,
$\vec{Z}_b$, can be changed during the flight time of two wave packets
which are short enough to distinguish times $t_a$, $t_b$.  The
detectors are widely separated, so the collisions of the wave packets
with the detectors lie outside the light cones of each other.
Experiments have strongly confirmed the absence of local hidden
parameters, since Bell's inequalities [10] are violated.  It is not
clear how to check experimentally the existence or nonexistence of
nonlocal physical influence [1], a problem connected with the
EPR ``paradox'' [11].  However, we will show here that there is no
need to invoke nonlocal influence to explain the results of EPR
experiments.

In [8] the authors used elastic scattering of very slow protons, $p +  p
\rightarrow p + p$ in order to get the singlet state of a $pp$ pair
(that is, before interaction with the detectors):
$$
\mid \psi > = \frac{1}{\sqrt{2}} \left\{ \mid \uparrow (\vec{Z})
\vec{R}_a t_a > \mid \downarrow (\vec{Z}) \vec{R}_b t_b > - \mid
\downarrow (\vec{Z}) \vec{R}_a t_a > \mid \uparrow (Z) \vec{R}_b t_b >
\right\}~, 
\eqno(66)
$$

Since $\mid \psi >$ in this case is a rotation invariant, axis
$\vec{Z}$ in the superposition can be directed arbitrarily.

\noindent{\bf Question 1.}  (66) is a macroscopically nonlocal state.
$\vec{R}_a$, $\vec{R}_b$ are widely separated but every particle
is ``present'' in {\underline{both}} branches.  What does this mean?

The answer, according to Secs.~1 and 2 is that wave functions contain 
information about physical systems.  Although particles are more or less
located, information about them can be nonlocal.  {\it
Information, and not a particle, is present simultaneously in both
branches.} 

\noindent{\bf Question 2.}  If $\mid \psi >$ is a container of
information, then it 
can be described verbally.  But how?

A language which can be {\it explicitly} used to describe a quantum 
state can be defined by a measuring apparatus corresponding exactly
to that state.  In case (66) this  can be an apparatus 
measuring observable $\Sigma$,
$$
\hat{\Sigma} = \frac{1}{2} \left( 1 + \vec{\sigma}_{a}
\vec{\sigma}_{b} \right)~~,
\eqno(67)
$$
\noindent where $\sigma_x$, $\sigma_y$, $\sigma_z$ are Pauli matrices.
Such an apparatus would have two pointer positions, $+1$ for the triplet
$(pp)$ state, and $-1$ for the singlet $(pp)$ state.  The latter is our
$\mid \psi >$ state:
$$
\hat{\Sigma} \mid \psi > = - \mid \psi >~~.   
\eqno(68)
$$
\noindent $(+1, -1)$ form the language corresponding to such an
apparatus. ``$-1$'' explicitly expresses $\mid \psi >$, and nothing
forbids us to 
describe $\mid \psi >$ in terms of that apparatus (i.e., in the
$\Sigma$-representation).   However, we cannot
express $\mid \psi >$ explicitly  in the language corresponding to the
apparatus of Fig.~5, which is measuring $\sigma_{Z_a} \cdot
\sigma_{Z_b}$, because $\sigma_{Z_a} \sigma_{Z_b}$ does not commute
with $\hat{\Sigma}$.  Therefore, there is no computable translation
between the truth values of $\Lambda_\psi$ and the truth values of
statements describing $\sigma_{Z_a} \sigma_{Z_b}$.

\noindent{\bf Question 3.}   According to (66), the spin of particle
$a$ is completely {\underline{uncertain}}.  However, by measuring
$S_{Z_a}$ we get a certain $S_{Z_a}$-projection along a certain axis.
How does this happen?

The answer is a corollary of the answer to the previous question.
Since there are no computable functions translating language
explicitly corresponding to physical state $\mid \psi >$ into 
language explicitly corresponding to the pointer positions of the
$d_a$, $d_b$ detectors, there are no logical connections between state
$\mid \psi >$ and the result of its measurements by the apparatus of
Fig.~5.  (Logical connections mean, in particular, the existence of an
algorithm connecting the states before and after the
measurement.)  Therefore, every individual measurement in this case gives us
completely {\underline{new}} information, unconnected with the
previous information.  The new information is contained in the new state
created by the measurement; it can, for example, be $\mid \uparrow (\vec{Z}_a)
\vec{R}_a t > \mid \downarrow (\vec{Z}_b) \vec{R}_b t >$, if the
result of the measurement is $S_a = \frac{1}{2}$, $S_b = -
\frac{1}{2}$.  The old information is destroyed at that moment 
when, and at the place where, the logical system processing the
measurement has formulated the result of the individual measurement.

\noindent{\bf Question 4.}  When measuring $S_{Z_a}$ in the 
$a$-channel, we have chosen axis $Z_a$  completely arbitrarily.
How does particle $b$ know that its spin must be directed against axis
$\vec{Z}_a$? 

The answer is that it does not.  All information about the 
logical correlations between the two branches, $a$ and $b$, of the common
state $\mid \psi >$, was created simultaneously with this state, i.e.,
well before the measurement.

Let us calculate the correlation mentioned in the question, using the
results of Sec.~2.

Let statement $M_a (\vec{Z}_a)$ be
$$
M_a \left( \vec{Z}_a \right) = {\mbox{``}} S_{Z_a} (a) = + \frac{1}{2}
{\mbox{'';}}~ M_a(\vec{Z}_a) \Rightarrow \hat{M}_a ( \vec{Z}_a) =
\mid \uparrow (\vec{Z}_a) \vec{R}_a t_a >< \uparrow (\vec{Z}_a)
\vec{R}_a t_a \mid~~,
\eqno(69)
$$
\noindent and statement $M_b (\vec{Z}_b)$ be
$$
M_b \left( \vec{Z}_b \right) = {\mbox{``}} S_{Z_b} (b) = - \frac{1}{2}
{\mbox{'';}}~ M_b (\vec{Z}_b) \Rightarrow \hat{M}_b ( \vec{Z}_b) =
\mid \downarrow (\vec{Z}_b) \vec{R}_b t_b >< \downarrow (\vec{Z}_b)
\vec{R}_b t_b \mid~~. 
\eqno(70)
$$
Here axes $\vec{Z}_a$ and $\vec{Z}_b$ can be different. 

If $\vec{Z}_a = \vec{Z}_b = \vec{Z}$, then the logical equivalence,
$$
M_a \left( \vec{Z} \right) \longleftrightarrow M_b \left( \vec{Z} \right)~~,
\eqno(71)
$$
\noindent is true in state $\mid \psi >$ throughout the history of the
particles' movement.  Indeed, from (30), (66), (69), (70),
$$
M=M_a \longleftrightarrow M_b \Rightarrow \hat{M} = \hat{1} - \hat{M}_a -
\hat{M}_b +  2 \hat{M}_a \hat{M}_b~;~~ Tr \left( \Lambda_\psi \mid M
\right) = < \psi \mid \hat{M} \mid \psi > =  1~~.
\eqno(72)
$$
\noindent Thus, if spin $S(a)$ is up (down), then spin $S(b)$ is down
(up).  (In the case of 1/2 spin, the negations of the
statements ``$S_Z = \pm 1/2$'' are the statements ``$S_Z = \mp
1/2$''.)  It is extremely important to note that 
statement (71) does not specify whether any spin is
really directed along or against any given axis.  It is a purely logical
assumption. 

The existence of a logical correlation (71) between $a$ and $b$ branches
does not mean that there is any nonlocal mutual influence; in quantum
mechanics, according to Sec.~1, almost all wave
functions~---~the containers of information~---~ are ``nonlocal''.

The logical correlation between $M_a (\vec{Z}_a)$ and $M_b (\vec{Z}_b)$
in state $\mid \psi >$, when $\vec{Z}_b \neq \vec{Z}_a = \vec{Z}$, is
equal to the truth of $M_a \wedge M_b$.  Using $< \psi \mid \hat{M}_a
\mid \psi > = < \psi \mid \hat{M}_b \mid  \psi > = \frac{1}{2}$, we
get 
$$
\begin{array}{rcl}
Tr \left( \Lambda_\psi \mid M_a \wedge M_b \right) & = & < \psi \mid
\hat{M}_a \hat{M}_b \mid \psi > = \mid < \uparrow (\vec{Z}) \vec{R}_b
t_b \mid \uparrow ( \vec{Z}_b) \vec{R}_b t_b > \mid^2 = \\
& = & \frac{1}{2} \cos^2 \theta/2~~,
\end{array}
\eqno(73)
$$
\noindent and
$$
Tr (\Lambda_\psi \mid M_a \leftrightarrow  M_b) = \cos^2 \theta/2~~,
\eqno(74)
$$
\noindent (see Fig. 5).  The result does not depend on the order of
the cofactors since $M_a$ and $M_b$ describe different channels, and 
therefore commute.

\noindent{\bf Question  5.} Let axes $\vec{Z}_a$ and $\vec{Z}_b$ be
not parallel.  We will measure simultaneously $S_{Z_a} (a)$ and
$S_{Z_b} (b)$.  Suppose the results are $S_{Z_a} (a) = + \frac{1}{2}$
and $S_{Z_b} (b) = - \frac{1}{2}$.  According to (69), (70), (71) and
(72), if $S_{Z_a} (a) = + \frac{1}{2}$, then $S_{Z_a} (b) = -
\frac{1}{2}$.  ($S_{Z_a} (b)$ means that, in channel $b$, $\vec{Z}_b =
\vec{Z}_a$.)  However, we have measured $S_{Z_b} (b) = - \frac{1}{2}$.
This means that we have measured simultaneously the noncommuting
$S$-projections along two non-parallel axes.  Is this permitted by
quantum mechanics?

The answer is that nothing in quantum mechanics forbids us to measure
two noncommuting observables at {\underline{different}} times and/or
places.  The logical correlation between the two results, $S_{Z_a} (a) = +
\frac{1}{2}$ and $S_{Z_b} (b) = - \frac{1}{2}$, see formula (73),
existed before our measurement.  Since the theory is correct, the
measurement should confirm this formula.  Note that when (73) is an
evaluation of truth, it needs only logical confirmation, not an 
experimental one.  When we use (73) to evaluate probability,
we need the measurement on an ensemble (since $\frac{1}{2} \cos^2
\theta/2 < 1$, and $\theta \neq 0$).

\noindent{\bf Question 6.}  (A formulation of the EPR
``paradox''.) 

Let consecutive measurements of $S_Z$-projections of two different
particles onto two different axes, $\vec{Z}_a$ and $\vec{Z}^\prime_a$,
in the same $a$-channel result in $S_{Z_a} (a) = \frac{1}{2}$ and
$S_{Z^\prime_a} (a) = \frac{1}{2}$.  In the first case we predict that
$S_{Z_a} (b) = - \frac{1}{2}$, in the second case that $S_{Z^\prime_a}
(b) = - \frac{1}{2}$.  
Therefore, we predict with certainty the values of $S_{Z_a}(b)$ and
$S_{Z^\prime_a}(b)$, represented by {\underline{noncommuting}} operators
without any interaction with the particle in the $b$-channel.  As
formulated in [11], 
``If without in any way disturbing a system, we can predict with
certainty (i.e., with probability equal to unity), the value of a
physical quantity, then there exists an element of physical reality
corresponding to this physical quantity.''  Which two elements of
physical reality in channel-$b$ correspond to the seeming appearance
there of two numerical values of two noncommuting observables (at
different times)?

The answer is that there are no such elements of physical reality in
the $b$-channel.  The information about all logical correlations
between the two channels is the only physical reality available to
science; and this reality, information, is nonlocal in the sense
explained in Sec.~4 and existed before the measurements verifying
logical correlations.  In this case, we  have two correlations (71)
with two different axes $\vec{Z}$.  They belong to the same vector
(66) but to different pairs of particles.  This assumes that wave
functions describe {\underline{individual}} systems, not ensembles
(although ensembles of identical systems can be prepared). 

Indeed, wave functions are collapsed and
created only individually; there is therefore no reason to interpret
them as describing only ensembles of identical systems.  Moreover,
since conditional truths of statements about a system, whose wave
function is a packet depend on the shape of the packet, and this
shape may be unique, we can apply wave functions to individual
systems first, and then to ensembles if ensembles is prepared.

\pagebreak

\centerline{\bf Conclusion}

\bigskip

The mysteries of quantum mechanics can be understood if we recognize
that at the quantum level of accuracy we enter a world in which logic
and language become parts of nature, and where we have to deal with
noncomputable functions.  However, full understanding of these
mysteries will be reached only when a similar approach is developed to
quantum field theory.

In a radically different direction, this understanding of quantum
phenomena can be useful in developing a theory of mind.  After all,
quantum nonlocality is nonlocality of logic and language.

\bigskip\bigskip
\bigskip

\begin{flushright} July 15, 1996
\end{flushright}

\bigskip\bigskip
\bigskip
\bigskip\bigskip
\bigskip
\bigskip\bigskip
\bigskip
\bigskip\bigskip
\bigskip

\centerline{\bf Acknowledgments}

Various parts of this paper have been presented over the past two
years at seminars and colloquia at Cornell and elsewhere, most
recently at the Physics Division colloquia at Berkeley.   I would like
to thank participants in these and other discussions, especially Kurt
Gottfried, David Mermin, Henry Stapp, and Philippe Eberhardt, for
their constructive comments.

\pagebreak

\centerline{\bf References}

\bigskip

\begin{enumerate}

\item Quantum Mechanics versus Local Realism.  The
Einstein-Podolsky-Rosen Paradox.  Edited by Franco Selleri.  Plenum
Press, 1988. 

\item Y. F. Orlov.  The Logical Origins of Quantum Mechanics.  Annals
of Physics, vol. 234, No. 2, Sept. 1994.

\item J. von Neumann.  Mathematische Grundlagen der Quanten-Mechanik.
Springer-Verlag, 1932.

\item S. C. Kleene.  Introduction to Metamathematics.
Wolters-Noordhoff and North-Holland, 1971.

\item W. Gerlach and O. Z. Stern, Z. Phys. {\bf 9}, 349 (1922).

\item Kurt Gottfried.  Quantum Mechanics. Addison-Wesley, 1966.

\item Martin Davis, Computability and Unsolvability.
Dover, 1982. 

\item Kurt G\"odel, Monatshefte f\"ur Mathematik und Physik, vol. 38,
pp. 173-198 (1931).

\item M. Lamehi-Rachti and W. Mittig.  Phys. Rev. D. {\bf 14}, No. 10,
2543 (1976).

\item J. S. Bell.  Review of Modern Physics, {\bf 38}, 447 (1966).

\item  A. Einstein, N. Podolsky and B. Rosen.  Phys. Rev. {\bf 47},
777 (1935).

\end{enumerate}

\pagebreak

\epsfysize = 3in
\centerline{\epsfbox{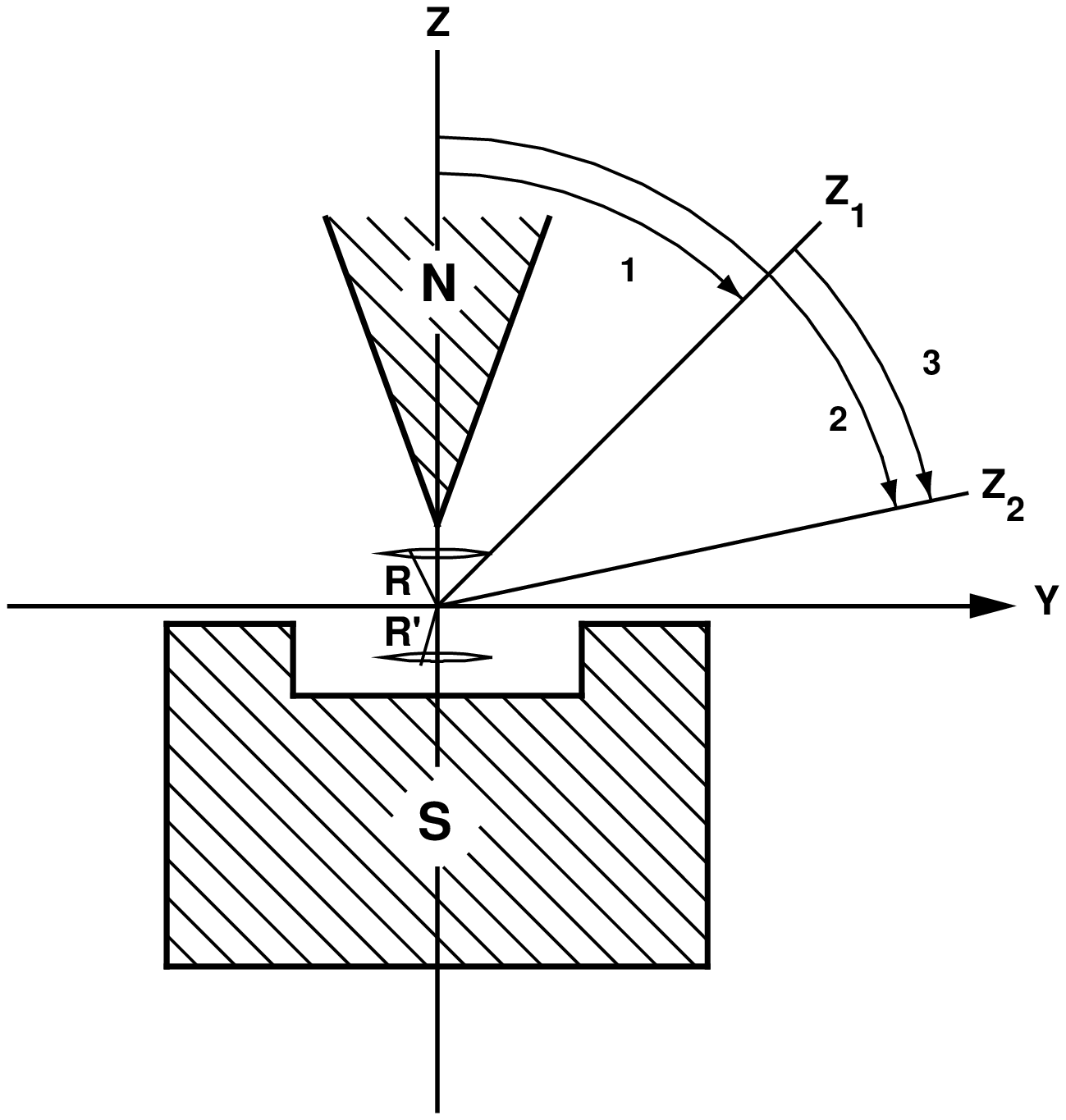}}
\centerline{Fig. 1}

\vspace{1.5in}
\vspace{1.5in}
\centerline{\fbox{See external graphics file.}}
\vspace{1.5in}

\centerline{Fig. 2}

\pagebreak

\vspace*{0.5in}
\epsfxsize = 4in
\centerline{\epsfbox{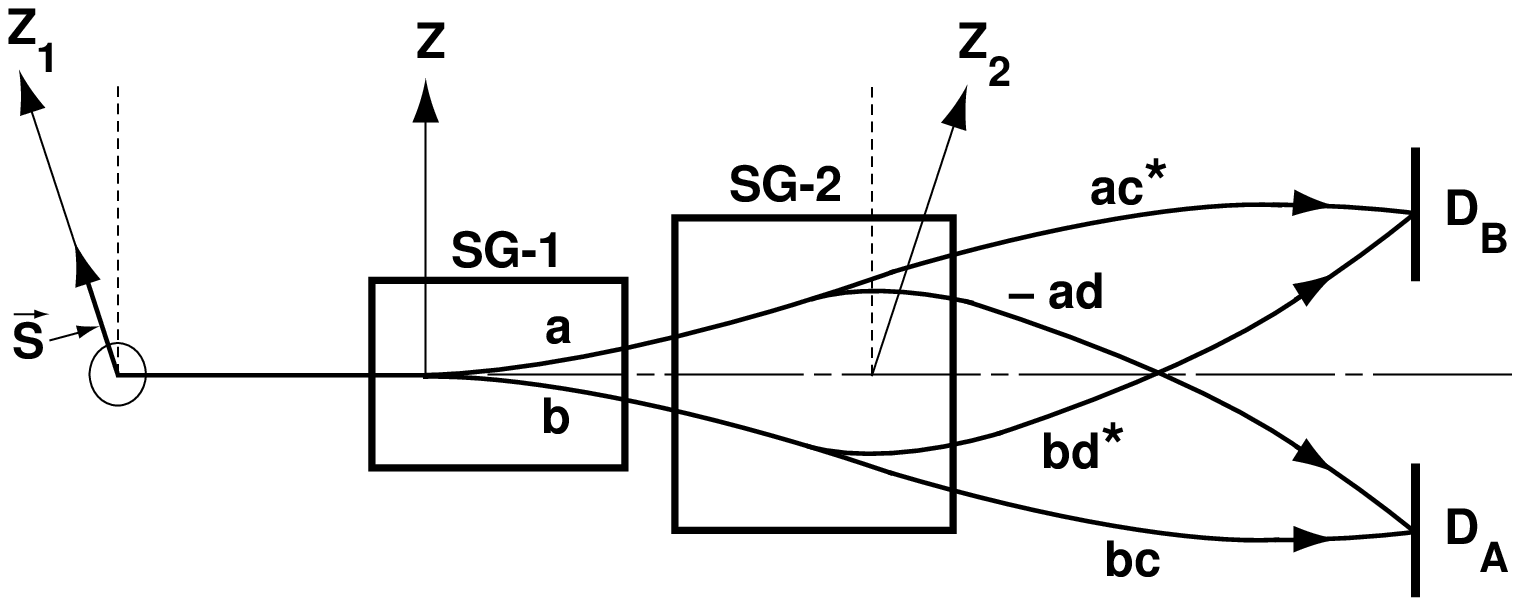}}

\centerline{Fig. 3}

\vspace{1in}

\epsfysize = 3in
\centerline{\epsfbox{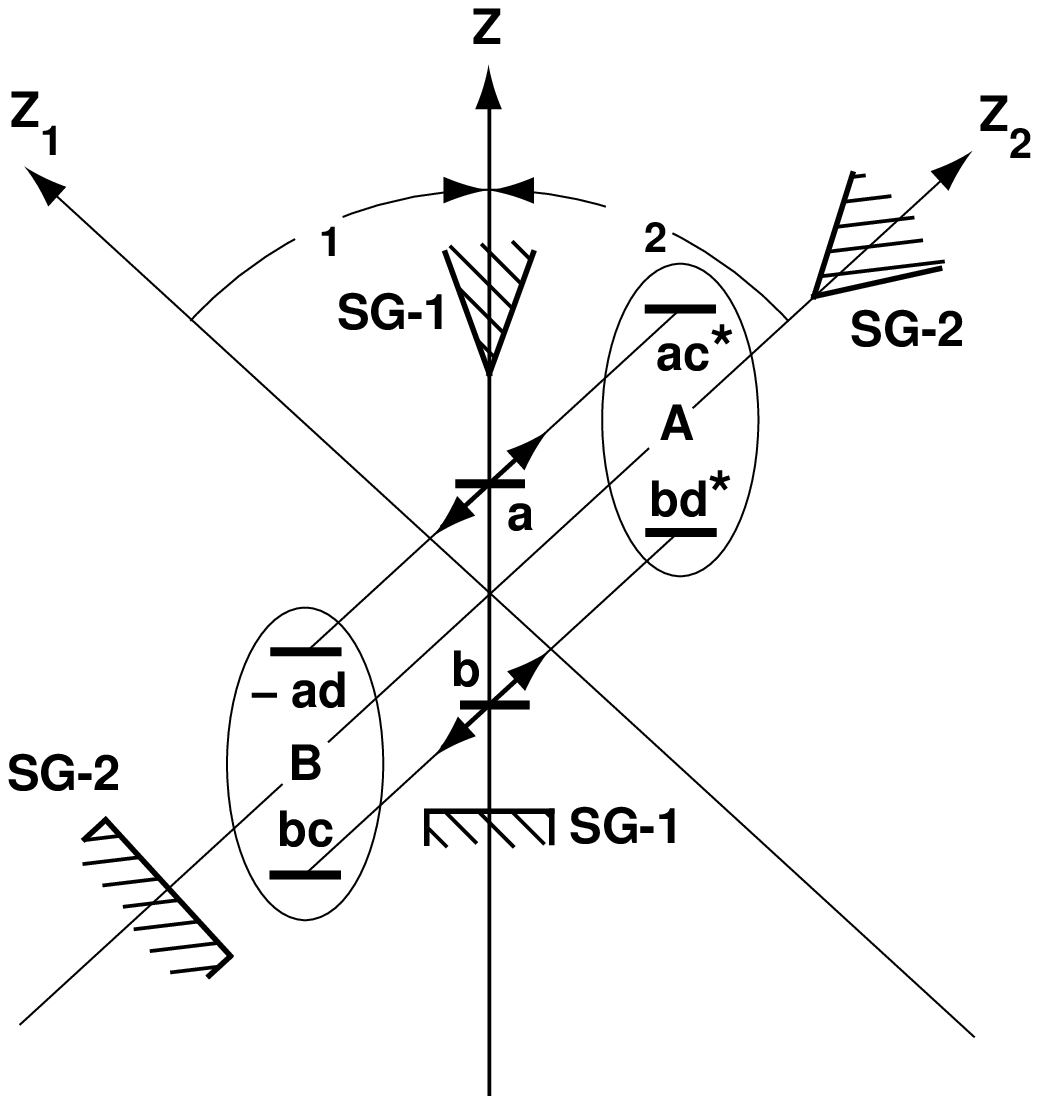}}

\centerline{Fig. 4}

\pagebreak

\epsfxsize = 5.5in
\centerline{\epsfbox{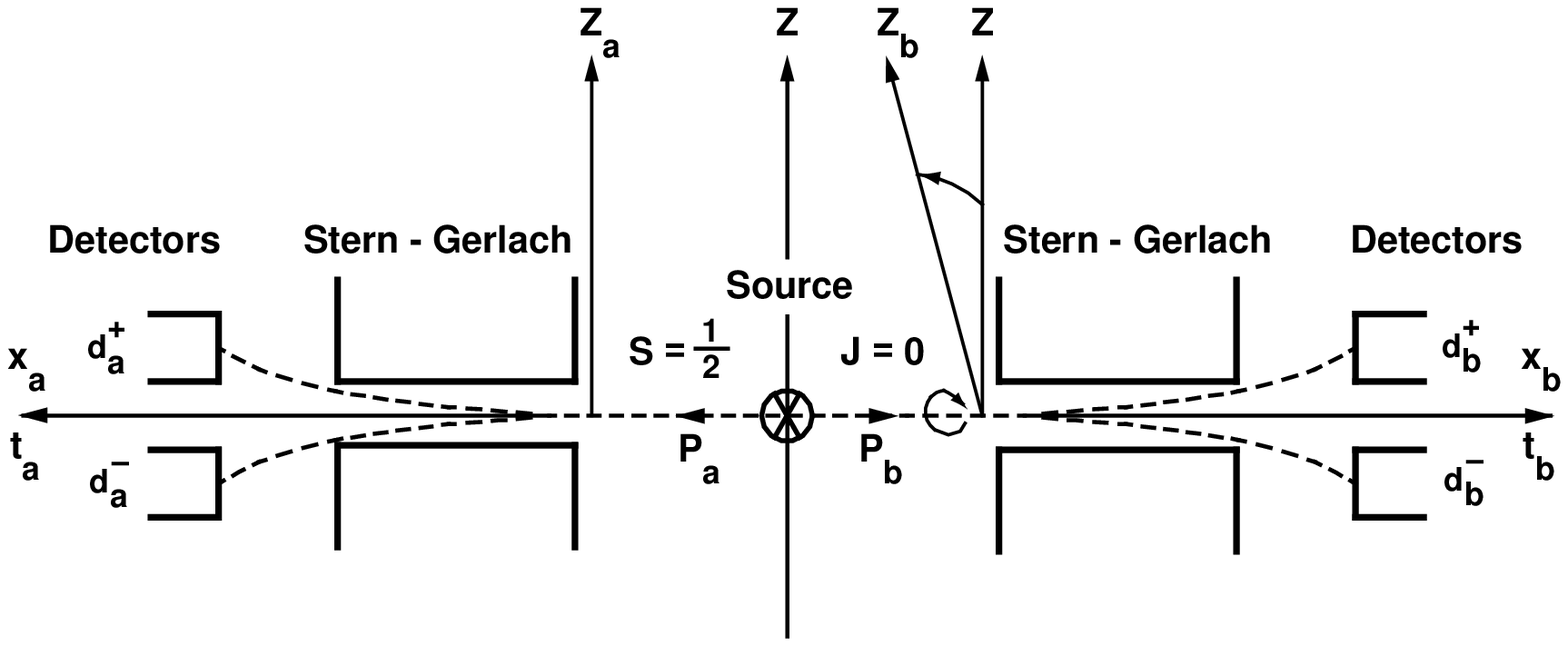}}

\centerline{Fig. 5}

\end{document}